\documentclass[aps,prb,twocolumn,amsmath,amssymb,reprint,superscriptaddress,longbibliography]{revtex4-1}

\usepackage[utf8]{inputenc}
\usepackage[english]{babel}
\usepackage{amsmath}
\usepackage{graphicx}
\usepackage[dvipsnames]{xcolor}
\usepackage{dcolumn}
\usepackage{bm}
\usepackage{makeidx}
\usepackage[alsoload=synchem]{siunitx} 
\usepackage{physics}
\usepackage[colorlinks, citecolor={blue!50!black}, urlcolor={blue!50!black}]{hyperref}
\usepackage{cleveref}	

\begin{document}
\title{Designing three-dimensional flat bands in nodal-line semimetals}

\author{Alexander Lau}
\thanks{These two authors contributed equally}
\affiliation{International Research Centre MagTop, Institute of Physics, Polish Academy of Sciences, Al. Lotnik\'ow 32/46, 02-668 Warsaw, Poland}

\author{Timo Hyart}
\thanks{These two authors contributed equally}
\affiliation{International Research Centre MagTop, Institute of Physics, Polish Academy of Sciences, Al. Lotnik\'ow 32/46, 02-668 Warsaw, Poland}
\affiliation{Department of Applied Physics, Aalto University, 00076 Aalto, Espoo, Finland}

\author{Carmine Autieri}
\affiliation{International Research Centre MagTop, Institute of Physics, Polish Academy of Sciences, Al. Lotnik\'ow 32/46, 02-668 Warsaw, Poland}

\author{Anffany Chen}
\affiliation{Department of Physics and Astronomy, University of British Columbia, Vancouver, BC, Canada V6T 1Z1}
\affiliation{Quantum Matter Institute, University of British Columbia, Vancouver BC, Canada V6T 1Z4}

\author{Dmitry I. Pikulin}
\affiliation{Station Q, Microsoft Corporation, Santa Barbara, California
93106-6105, USA}
\affiliation{Microsoft Quantum, Redmond, Washington 98052, USA}

\date{\today}

\begin{abstract}
Electrons with large kinetic energy have a superconducting instability for infinitesimal attractive interactions.
Quenching the kinetic energy and creating a flat band renders an infinitesimal repulsive interaction the relevant perturbation.
Thus, flat band systems are an ideal platform to study the competition of superconductivity and magnetism and their possible coexistence.
	Recent advances in the field of twisted bilayer graphene highlight this in the context of two-dimensional materials.
	Two dimensions, however, put severe restrictions on the stability of the low-temperature phases due to enhanced fluctuations.
	Only three-dimensional flat bands can solve the conundrum of combining the exotic flat-band phases with stable order existing at high temperatures.
	Here, we present a way to generate such flat bands through strain engineering in topological nodal-line semimetals.
	We present analytical and numerical evidence for this scenario and study the competition of the arising superconducting and magnetic orders as a function of externally controlled parameters.
	We show that the order parameter is rigid because the three-dimensional quantum geometry of the Bloch wave functions leads to a large superfluid stiffness in all three directions. Using density-functional theory and numerical tight-binding calculations we further apply our theory to strained rhombohedral graphite and CaAgP materials.
\end{abstract}

\maketitle

\section{Introduction}
The study of correlated many-particle states in flat-band systems goes back to the consideration of few-particle nuclear-physics systems in the 1960s when S.~T.~Belyaev demonstrated that, in the presence of degenerate single-particle states, interactions can lead to a pairing gap increasing linearly with the interaction strength~\cite{Belyaev}.
Since then, there has been a fruitful exchange of ideas between the nuclear-physics and the condensed-matter communities exploring analogies between nuclear-physics systems and ultrasmall superconducting grains \cite{RICHARDSON1964221, RevModPhys.76.643, VONDELFT200161}.
Recently, two-dimensional (2D) flat bands have provided new ground for exotic states of condensed matter~\cite{jain_2007, Moon1995, Kopnin2011,  Annica2017, Ojajarvi2018}. 
In particular, the advances in the fabrication of flat bands in twisted bilayer graphene have attracted a lot of attention due to the novel exotic phases becoming available experimentally~\cite{cao2018unconventional, cao2018correlated, Lu2019, Serlin2020, Chen2020, stepanov2019interplay, saito2020decoupling}.
Other setups and materials realizing 2D flat bands have been studied~\cite{Lee2016,Yin2019,ma2020spinorbitinduced} but, to our knowledge, their three-dimensional (3D) counterparts~\cite{khodel1990, fu2020} have not been explored in realistic materials.
In the present manuscript, we carry this development to its logical end point by proposing a viable approach for realizing 3D flat-band systems.

We combine two actively-studied ingredients to manifest a 3D flat band: nodal-line semimetals (NLSMs)~\cite{chiu2016classification} and strain engineering in topological semimetals~\cite{ilan2019pseudo}.
NLSMs are materials where topologically protected band crossings form a line (nodal line) in the Brillouin zone, provided that certain symmetries are satisfied.
Additionally, topologically protected drumhead surface states appear inside the region bounded by the projection of the nodal line onto the 2D surface Brillouin zone.
The drumhead surface states exhibit a nearly flat dispersion if the material has approximate chiral symmetry, for example due to a sublattice structure~\cite{Kopnin2011}.

Strain engineering produces opportunities to generate Landau level-like flat bands in the absence of external magnetic fields~\cite{levy2010strain, TangFu14, amorim2016novel, Kauppila16, peri2019axial, Kim2019, manesco2020correlations}.
This is because the action of the strain near the Fermi surface resembles that of a magnetic field in the local area of the Brillouin zone.
In the case of Weyl/Dirac semimetals, there are simply two Fermi pockets near the Fermi energy and the action of strain near the pockets can be described as electromagnetic fields ~\cite{bevan1997momentum, liu2013chiral, shapourian2015viscoelastic}.
Here, we show that such properties of strain are more generic and can be applied to materials with nodal lines as well.
For stationary strain, the resulting pseudo-magnetic field then depends on the position along the nodal line.
There is, however, always one commonality in nodal systems insensitive to the strength and direction of the pseudo-magnetic field -- the zeroth pseudo-Landau level (PLL).
Therefore, one would expect that the zeroth PLL forms a \textit{3D flat band}, while the higher PLLs are flat only in two dimensions and have a dispersion along the nodal line.

We confirm this intuitive argument above using numerical and analytical arguments.
We show that the zeroth PLL indeed forms a 3D flat band, which evolves from the drumhead surface states of the NLSM, and obtain wavefunctions thereof.
Using these wavefunctions and assuming competition between magnetism and superconductivity, we obtain the phase diagram of the system as a function of the filling factor and the interaction strengths. We show this system is a promising platform for studying intertwined phases~\cite{Fradkin2015,Ojajarvi2018} as it can be tuned \emph{in-situ} from the magnetic to the superconducting phase by controlling the magnitude of the strain.

\begin{figure*}
	\includegraphics[width=\linewidth]{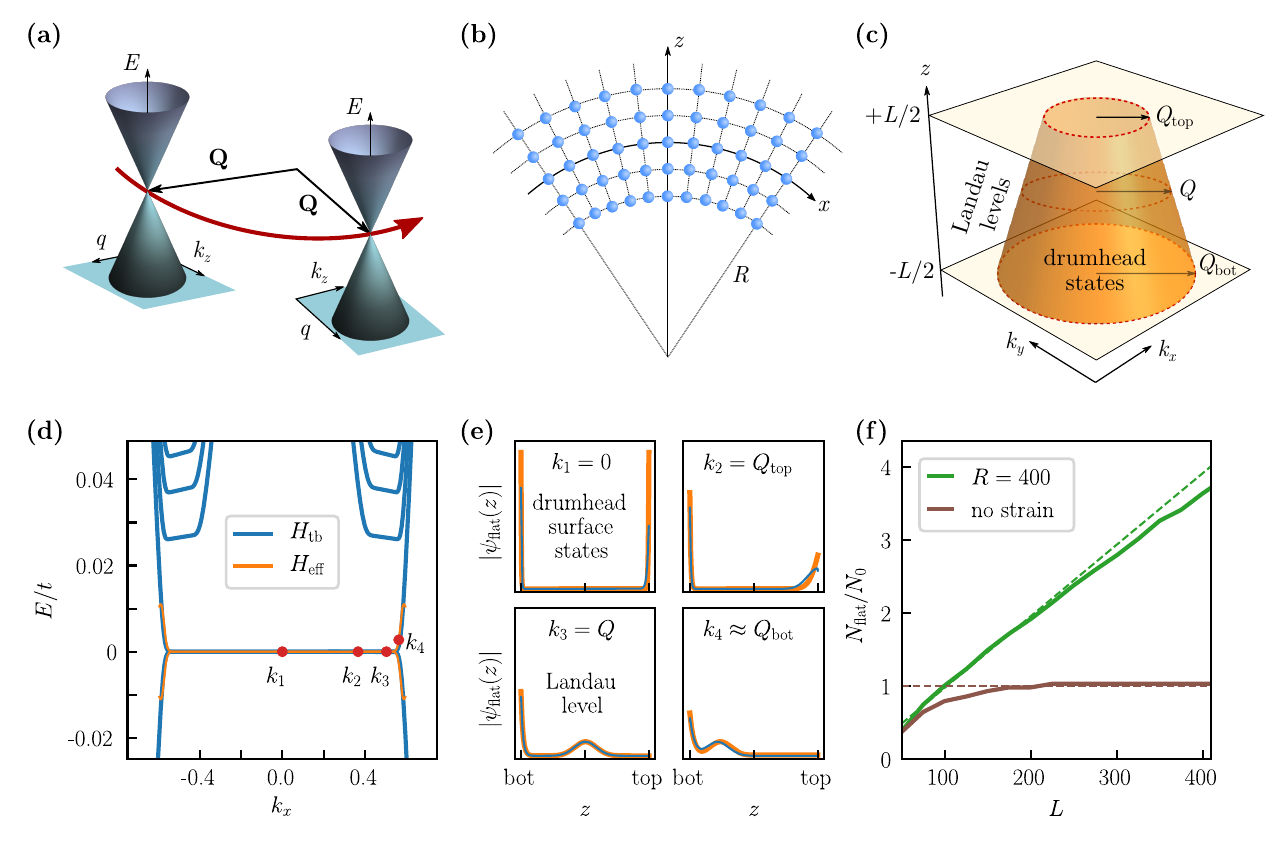}
	\caption{(a) Schematic representation of a nodal-line semimetal bandstructure. Here, $\mathbf{Q}$ is a vector pointing to a momentum on the nodal line such that $|\mathbf{Q}|=Q$ is the radius of a circular nodal loop. A Dirac cone is formed with respect to the perpendicular momentum components $q$ and $k_z$. (b) Example of a strain profile leading to the formation of a 3D flat band. The strain can be created e.g.~by bending the sample or growing it on a cylindrical surface. $R$ is the radius of the cylinder in the middle of the sample. (c) Bulk-boundary correspondence of the zeroth PLL of the NLSM. Due to the strain, the radius of the nodal circle varies as a function of $z$, such that the radius of the momentum-space area of the drumhead surface states on the top surface $Q_{\rm top}$ is different from the bottom surface $Q_{\rm bot}$. Zeroth PLL bulk states appear in the momentum space region between $Q_{\rm top}$ and $Q_{\rm bot}$. (d) Spectrum of the model Hamiltonian $H_\mathrm{tb}$ [Eq.~(\ref{eq:Hamiltonian})] in the presence of strain [Eq.~(\ref{eq:pseudo_magnetic_field})] with parameters $t_1=0.25t$, $t_2=0.8t$, $L=1000$, and $R=8000$. For comparison, we also show the energies of the effective Hamiltonian $H_\mathrm{eff}$ (orange bands) constructed from the analytical low-energy solutions (see Appendix below). (e) Numerical (blue) and analytical (orange) wavefunctions of the zeroth PLL at different momenta $k_x=k_i$ ($i=1,2,3,4$) indicated in (d). The drumhead surface states appear at both surfaces for $|k_x| < Q_{\rm top}, Q_{\rm bot}$. At $|k_x|=Q_{\rm top}$ the drumhead surface states from the top surface are deformed into the zeroth PLL bulk states and move towards the bottom surface with increasing $Q_{\rm top} < |k_x| < Q_{\rm bot}$. At $|k_x|=Q_{\rm bot}$ the PLL bulk state hybridizes with the drumhead surface state at the bottom surface. (f) Degeneracy of the flat band as a function of the height $L$ of the sample. Numerically, the degeneracy is calculated by integrating the density of states over an energy window ($|E|<10^{-3} t$). The model parameters $t_{1,2}$ are the same as in (d). Bold lines are the numerical results and dashed lines correspond to the analytical formulas. In the presence of strain, the degeneracy grows linearly with $L$ demonstrating the existence of a 3D flat band with degeneracy $N_{\rm flat}$ proportional to the volume of the sample (green dashed line). In the absence of strain, the degeneracy of the zero-energy drumhead surface states saturates with increasing $L$ demonstrating that it is proportional to the area of the surface of the sample, $N_0$.}
	\label{fig:fig1}
\end{figure*}

Moreover, we go beyond the mean-field picture by analyzing the properties of the collective modes and argue that the emerging order is more stable in the 3D case than in the previously studied 2D cases. Importantly, we show that the system supports a hitherto unexplored 3D quantum geometry of the Bloch wave functions that leads to a large superfluid stiffness in all directions despite the flatness of the bands. The 3D quantum geometry is a nontrivial consequence of the momentum dependence of the direction of the pseudo-magnetic field along the nodal line. Finally, we study the properties of 3D flat bands for the material examples of NLSMs belonging to CaAgP material class and of rhombohedral graphite.

\section{Non-interacting model}

\subsection{Two-band NLSM Hamiltonian and drumhead surface states}
A NLSM is a material for which the energy gap between two bands near the Fermi level closes along a line in the Brillouin zone [Fig.~\ref{fig:fig1}(a)].
The minimal model for a NLSM, believed to be applicable, for instance, to the CaAgP materials class (see below) and Ca$_3$P$_2$~\cite{xie2015new, chan20163, wang2017topological}, reads:
\begin{equation}
H(\boldsymbol{k})=\sigma^{z}\big(6t-t_{1}-2t \sum_{i=x, y, z} \cos k_{i} \big)+2t_{2}\sigma^{x}\sin k_{z}. \label{eq:Hamiltonian}
\end{equation}
Here $\sigma^{x, y}$ are Pauli matrices, and $t$, $t_1$, and $t_2$ are parameters determining the size of the nodal line and the Fermi velocity. Throughout this paper, the length scales are given in units of the lattice constant and the momentum is measured in units of the inverse lattice constant.
For $t_1\ll t$, this Hamiltonian has a gap closing along a circular path in the Brillouin zone -- a nodal loop -- given by $k_z=0$ and $(k_x, k_y)=\mathbf{Q}=Q(\cos\theta, \sin\theta)$, where $Q=\sqrt{t_1/t}$ and $\theta$ is the azimuthal angle.
The crossing between the two bands is protected by $x-y$ plane reflection symmetry. 
The bands are $\sigma^z$ eigenstates in the $k_z=0$ plane corresponding to reflection eigenvalues $\pm 1$. Taking a different viewpoint, the nodal line can also be considered to be protected by the chiral symmetry $\sigma^y$ or the combination of time-reversal and inversion symmetries. For this Hamiltonian, the drumhead surface states, appearing at momenta $k<Q$, have zero energy due to chiral symmetry. In the limit of small nodal-line radii $Q$, the corresponding surface wave functions at momentum $\mathbf{k}=\mathbf{Q}+\mathbf{q}$ can be written as
\begin{equation}
\Psi_{S, \pm}\big(\mathbf{Q}, \mathbf{q} \big)=\frac{1}{\sqrt{2 L_xL_y l_S \sinh\frac{L}{l_S}}} \left(\begin{array}{c}
1\\
\pm i
\end{array}\right)  e^{i(\mathbf{Q}+\mathbf{q})\cdot \mathbf{r}} e^{\pm \frac{z}{l_S}}, \nonumber
\end{equation}
where $\mathbf{q}=q(\cos\theta, \sin\theta)$ and $q<0$ describes the deviation from the nodal line. 
Here, $\pm$ correspond to top and bottom surfaces, $-L/2<z<L/2$, $l_S=-t_2/(t Q q)$ is the localization length of the drumhead surface states, and $L_x$ ($L_y$) is the length of the system in the $x$ ($y$) direction along which we have applied periodic boundary conditions.

\subsection{Strain and pseudo-Landau levels}

A simple way to realize strain with a constant gradient of the strain field is bending~\cite{liu2017quantum, ilan2019pseudo}, as shown in Fig.~\ref{fig:fig1}b.
The bend can be achieved by a proper choice of the growth substrate or by mechanical bending of the device. 
Such strain creates a displacement field $\boldsymbol{u}=(xz/R,0,0)$ with corresponding strain tensor components $u_{11}=z/R$ and $u_{13}=u_{31}=x/(2R)$.
In the limit of small nodal-line radii, the strain induces a pseudo-magnetic field (see Appendix below)
\begin{equation}
\mathbf{B}_5= \frac{\hbar}{el_B^2}\left(\sin \theta,-\cos\theta,0\right), \label{eq:pseudo_magnetic_field}
\end{equation}
where $\mathbf{Q}=Q(\cos\theta, \sin \theta)$ is a momentum along the nodal line and $l_B=\sqrt{RQ}$ is the pseudo-magnetic length. In this case, the analytical solution for the zeroth PLL wave function is 
\begin{equation}
\Psi_{0}\big(\mathbf{Q}, \mathbf{q} \big)=\frac{1}{\sqrt{2 L_xL_y}} \left(\begin{array}{c}
1\\
i
\end{array}\right)  e^{i(\mathbf{Q}+\mathbf{q})\cdot \mathbf{r}} \left(\frac{1}{\pi l_z^2}\right)^{1/4}e^{-\xi_q^{2}/2}. \nonumber
\end{equation}
Here $\xi_q=(z+z_q)/l_z$, $z_q=q l_B^2$ and the pseudo-magnetic localization length in the $z$-direction is given by
\begin{equation}
l_z=\sqrt{\frac{t_2}{t Q}} l_B. \nonumber
\end{equation}
Alternatively, one can perform a gauge transformation such that the wave function is localized in the $x-y$ plane and a plane wave in the $z$-direction. In that case, the localization length is 
\begin{equation}
l_{xy}=\frac{l_B^2}{l_z}.\nonumber
\end{equation}
The prefactors in the localization lengths arise due to the elongations of the elliptical cyclotron orbits.

Going beyond the lowest-order expansion in $Q$ would lead to a $\theta$ dependence of the strength of the pseudo-magnetic field because the strain shrinks the nodal line anisotropically. Nevertheless, the energy of the zeroth PLL is independent of the strength of the pseudo-magnetic field and, hence, in all cases we robustly obtain the desired 3D flat band. Since the pseudo-magnetic field originates from the spatial variation of the radius of the nodal loop, it can also be created by varying the relative amount of P and As contents in the CaAgP$_{1-x}$As$_x$ alloys along the $z$-direction, for example, by changing the crystal growth conditions with time.

\subsection{Connection of drumhead surface states and pseudo-Landau level}

In Fig.~\ref{fig:fig1}(c, d), we plot the spectrum of the lowest PLL states. We find that the zeroth PLL is two-fold degenerate: the bulk states discussed above coexist with drumhead states localized to the bottom surface.
We further observe that the PLL bulk states evolve from drumhead states localized to the \emph{top} surface, similar to the connection between Fermi arcs and PLLs in Weyl semimetals~\cite{grushin2016inhomogeneous}.
By using approximate analytic solutions for the drumhead surface states and for the PLL bulk wave functions, and by considering their coupling at the surface of the sample, we obtain an effective low-energy Hamiltonian (see Appendix below) that accurately describes the exact numerical results as shown in Fig.~\ref{fig:fig1}(d, e).
We confirm the 3D nature of the flat band by plotting the scaling of the total number of states in the flat band for different thicknesses $L$ with and without strain in Fig.~\ref{fig:fig1}(f).
For sufficiently large $L$, the number of states for the drumhead surface states without strain saturates to the value $N_0=Q^2 L_x L_y/(2 \pi)$.
On the contrary, the number of states, including the PLL bulk states, in the presence of strain grows linearly with $L$ as $N_{\rm flat}=Q L_x L_y L/(2 \pi l_B^2)$ -- confirming our hypothesis of a flat band that is genuinely 3D.

The interaction effects depend on the $\mathbf{k}$-space extent of the flat bands. Unlike in twisted bilayer graphene, the flat band in these 3D systems covers only a fraction of the Brillouin zone. In twisted bilayer graphene, however, the area of the moir{\'e}-Brillouin zone is inversely proportional to the square of the superlattice lattice constant and the latter is large around the magic angle. Thus, we expect that interaction effects in the considered 3D flat band systems are as important as in the case of twisted bilayer graphene (see below).

\section{Effects of interactions}

In flat-band systems, the density of states diverges and, therefore, interaction effects are important. Moreover, these systems have instabilities both with respect to repulsive and to attractive interactions, so that competition between various symmetry-broken phases is expected to be a generic feature of flat-band systems. We now turn our attention to the effect of an attractive pairing interaction and Coulomb repulsion between electrons.

\subsection{Magnetism}

For the study of magnetism, we self-consistently solve the mean-field equations for the magnetic order parameter $m_z$ and the chemical potential $\mu$ in the presence of the density constraint that the filling factor of the band be fixed to $C$ (see Appendix below)  
\begin{eqnarray}
m_{z} &=& \frac{V_0}{2} \bigg[n_F(-m_z-\mu) - n_F(m_z-\mu) \bigg], \nonumber \\
C &=& n_F(-m_z-\mu) + n_F(m_z-\mu).\label{magnetism-main}
\end{eqnarray}
Here, $V_0$ is the effective interaction strength and $n_F$ is the Fermi function. The filling factor $C$ is restricted to $0 \leq C \leq 2$, such that $C=0$ corresponds to the situation where the flat-band states are completely empty. whereas for $C=2$ both spin-up and spin-down flat-band states are fully occupied. The effective interaction strength $V_0$ is computed by projecting the Coulomb interaction to the zeroth PLL. Furthermore, it can be tuned by varying the strain (see Appendix below).
The critical temperature for magnetism depends on $C$ as
\begin{equation}
k_B T_{c, m} = \frac{V_0}{4} C (2-C). \label{Tc-FM-main}
\end{equation}
For conservative model parameters we estimate $T_{c, m} = 3\,\mathrm{K}$ (see Appendix below).

We point out that this theory describes several different types of magnetic order parameters as their projections to the PLL bulk wave functions are the same (see Appendix below). However, the surface effects break the symmetry explicitly and distinguish the magnetic order parameters from each other. We find that the lowest energy state corresponds to the situation where the magnetic order is staggered with respect to the PLL bulk states and the drumhead states at the bottom surface (see Appendix below). We note that the exact nature of the magnetic order is not important for the discussion of competing phases below since we focus on the bulk states here.

\subsection{Superconductivity}

Starting from the reduced BCS Hamiltonian with the density constraint, the mean field equations for the superconducting order parameter $\Delta$ and the chemical potential $\mu$ are (see Appendix below)
\begin{eqnarray}
\Delta&=&G_{0} \frac{\Delta}{2 \sqrt{\mu^2+\Delta^2}} \tanh(\beta \sqrt{\Delta^2+\mu^2}/2), \nonumber\\
C&=&1+\frac{\mu}{\sqrt{\mu^2+\Delta^2}} \tanh(\beta \sqrt{\Delta^2+\mu^2}/2), \label{superconductivity-main}
\end{eqnarray}
where $G_0$ is the effective interaction strength and $\beta=1/(k_B T)$.
The critical temperature is given by
\begin{equation}
k_BT_{c, sc}=\frac{G_0}{4} \frac{C-1}{{\rm arctanh}(C-1)}. \label{Tc-SC-main}
\end{equation}
For typical model parameters  we estimate $T_{c,sc}=1\,\mathrm{K}$ (see Appendix below).

We note that this value is similar to the critical temperatures observed in twisted bilayer graphene~\cite{cao2018unconventional, cao2018correlated}.
In the case of 3D flat bands in NLSMs, however, the stronger suppression of order-parameter fluctuations and the greater variability of parameters with strain can potentially lead to even larger critical temperatures. The same applies to the magnetic phase described above.

\subsection{Competing phases}
\begin{figure}[t]
	\includegraphics[width=\columnwidth]{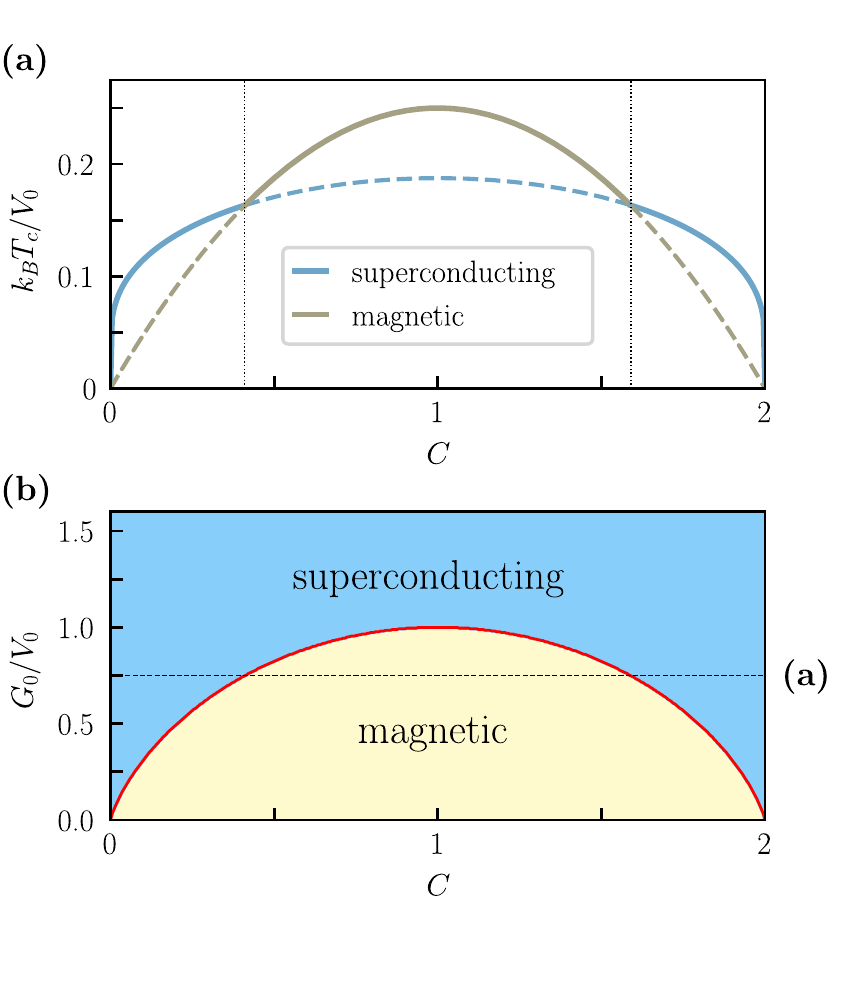}
	\caption{Phase diagram of the model Hamiltonian [Eq.~(\ref{eq:Hamiltonian})] in the presence of strain [Eq.~(\ref{eq:pseudo_magnetic_field})] and interactions [Eqs.~(\ref{magnetism-main}),(\ref{superconductivity-main})]. (a) Critical temperatures for magnetism [Eq.~(\ref{Tc-FM-main})] and superconductivity [Eq.~(\ref{Tc-SC-main})] as a function of the filling factor $C$ in the typical situation where the effective interaction strength for superconductivity is smaller than for magnetism  $G_0=0.75 V_0$. At half-filling $C=1$, the system realizes magnetic order but doping away from half-filling leads to superconductivity, resembling the phase diagrams experimentally observed in cuprates and twisted bilayer graphene. (b) Phase diagram as a function of $C$ and $G_0/V_0$ assuming that the system realizes the phase with the larger critical temperature. The dashed horizontal line represents a path through the phase diagram corresponding to (a).}
	\label{fig:fig2}
\end{figure}

As recent experiments in twisted bilayer graphene indicate, the competition between different correlated phases is a generic feature of flat-band systems~\cite{stepanov2019interplay, saito2020decoupling}. Moreover, in the case of competition between flat-band magnetism and superconductivity doping the system away from half-filling is generically expected to favor the superconducting phase~\cite{Annica2017}. Thus, we expect 
also magnetic and superconducting phases in 3D flat-band systems to be tunable using controllable parameters such as doping and the effective interaction strengths. The latter can be controlled, for example, by strain or an electrostatic environment. 

This expectation is confirmed by our calculations shown in Fig.~\ref{fig:fig2}. In Fig.~\ref{fig:fig2}(a), we plot the critical temperatures for magnetism [Eq.~(\ref{Tc-FM-main})] and superconductivity [Eq.~(\ref{Tc-SC-main})] as a function of the filling factor $C$. At half-filling $C=1$, the system realizes the phase with the larger effective interaction strength. Hence, for the typical situation where the Coulomb repulsion dominates the pairing interaction, $V_0>G_0$, the magnetic phase is realized. On the contrary, doping the system away from half-filling favors the superconducting phase. Therefore, flat-band systems are expected to have phase diagrams similar to those observed experimentally in cuprates~\cite{Lee06} and twisted bilayer graphene~\cite{cao2018unconventional, cao2018correlated}. In contrast to the complicated, strongly correlated Mott physics of cuprates~\cite{Lee06}, here the mechanisms of superconductivity and magnetism are independent of each other~\cite{Ojajarvi2018}. This is in agreement with experiments on twisted bilayer graphene~\cite{stepanov2019interplay, saito2020decoupling}. In addition to the filling factor $C$, the ground state can also be controlled by the relative interaction strength $G_0/V_0$, as shown in Fig.~\ref{fig:fig2}(b).

We note that our proposal offers a degree of control over $G_0$ and $V_0$ \textit{in a single sample}.
By varying the strength of the strain one changes the parameters as derived in the Appendix below, thus allowing to tune in situ between magnetic and superconducting phases.

We note that, in reality, the phase diagram may be more complicated due to possible co-existence of the order parameters near the mean-field phase boundary~\cite{Ojajarvi2018}. This may be analyzed using numerical techniques or functional renormalization group and are beyond the scope of the current manuscript. Nevertheless, we underline that 3D flat bands can shed light on the competition and intertwining of order parameters while avoiding the complications caused by strongly correlated Mott physics and the fluctuations present in low-dimensional systems.

\subsection{Superfluid stiffness}

To further substantiate our claims concerning the stability of the mean-field solutions, we analyze the collective modes of the system. In the case of 3D flat bands, the quasiparticle spectrum is fully gapped which means that the amplitude mode is gapped. Thus, the order parameter is stable against amplitude-mode fluctuations. This is in strong contrast to the order parameter appearing in the case of flat-band superconductivity due to 2D drumhead surface states, where the system is gapless and susceptible to strong amplitude-mode fluctuations~\cite{Kauppila2016}.

The phase rigidity, on the other hand, needs to be analyzed more carefully as kinetic contribution to it is negligible due to the flatness of the band. It is determined by the superfluid stiffness tensor $D_s$, which is related to the supercurrent $\mathbf{j}$ in the system as
\begin{equation}
j_i = \frac{2 e}{\hbar}\sum_j [D_s]_{ij}  \big(\partial_j \varphi-\frac{2e}{\hbar}A_j\big),
\end{equation}
where $\varphi$ is the phase of the superconducting order parameter.
The question whether a supercurrent even exists in the systems considered here is particularly relevant as the Fermi velocity within a featureless flat band is zero. Previous studies have shown that in 2D systems there can still exist a nonzero contribution to the superfluid stiffness caused by the quantum geometry of the Bloch wave functions~\cite{Moon1995, Pikulin2016, Peotta2015, Liang2017, Xiang2019, Xie2020, Julku2020, hu2020quantummetricenabled}.
However, the possibility of an analogous 3D quantum geometry that gives rise to superfluid stiffness in all directions of a 3D system has not been explored.
In the Appendix below, we combine methods developed in the context of quantum Hall systems~\cite{Moon1995, Pikulin2016} and flat-band superconductors~\cite{Peotta2015,  Xie2020} to calculate the superfluid stiffness for the model Hamiltonian [Eq.~(\ref{eq:Hamiltonian})] in the presence of strain [Eq.~(\ref{eq:pseudo_magnetic_field})].
We find that the superfluid stiffness perpendicular $D_{s,zz}$ and parallel $D_{s,xx}=D_{s,yy}$ to the layers have large geometric contributions
\begin{eqnarray}
D_{s,zz} =\frac{n_0 l_z^2}{2}  \Delta, \ D_{s,xx} =D_{s,yy} =   \Delta  \frac{n_0}{4} l_{xy}^2. \label{stiffness-main}
\end{eqnarray}
Therefore, the system has rigidity against phase fluctuations and supports large bulk supercurrents thereby confirming the 3D nature of the superconducting state. Here, $n_0=Q/(2 \pi l_B^2)$ is the degeneracy of the flat band per volume and the directional dependence of the
stiffness arises because of the elongations of the semiclassical cyclotron orbits.

We emphasize that the non-zero superfluid stiffness along all three directions, as obtained in this section, is not a property that flat bands exhibit by default, but the necessary 3D quantum geometry arises due to the variation of the pseudomagnetic field direction along the nodal line [Eq.~(\ref{eq:pseudo_magnetic_field}) and Appendix below].

\section{Material considerations}

\begin{figure*}[t]
	\includegraphics[width=\linewidth]{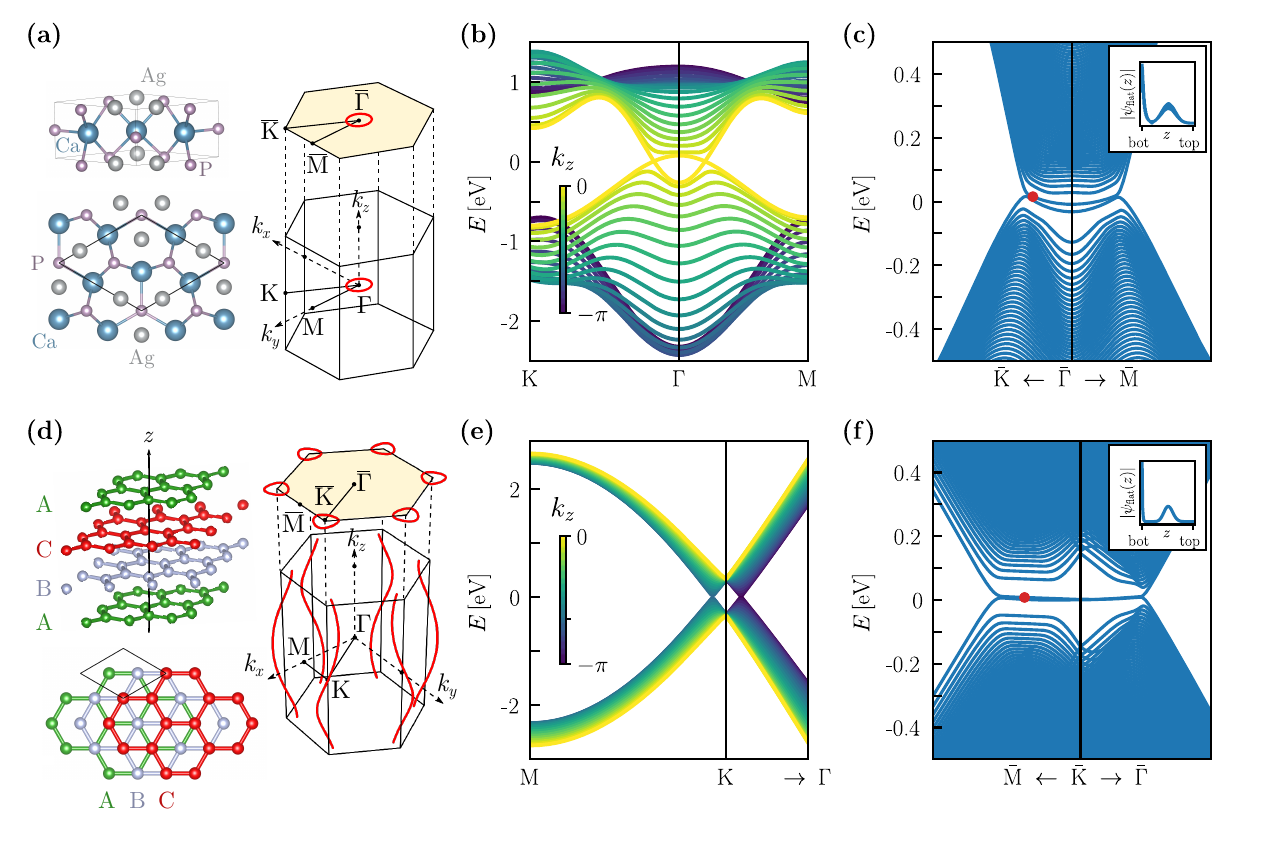}
	\caption{3D flat bands in nodal-line semimetals CaAgP (first row) and rhombohedral graphite (second row). (a) Top and side views of the CaAgP crystal structure. Next to these, we indicate the shape and position of its nodal circle (red), which is centered at the $\Gamma$ point of the hexagonal Brillouin zone. The beige hexagon represents the (001) surface Brillouin zone with the surface projection of the nodal loop. (b) Bulk bands of the two-band tight-binding model without strain along paths with fixed $k_z$. The bands only cross in the plane $k_z=0$. (c) Tight-binding spectrum of a strained (001) slab for $L=200$, $R=800$ with nearly flat bands. The inset shows the spatial profile of the flat-band wave function at the momentum indicated by the red dot. It is a superposition of a bottom-surface state and a (Gaussian) Landau-level state. (d) Top and side views of ABC-stacked (rhombohedral) graphite next to a sketch of its nodal lines. In contrast to CaAgP, there are several nodal lines that spiral around axes going through the K and K' points of the hexagonal Brillouin zone. Their projections into the surface Brillouin zone form approximate nodal circles. (e) Bulk bands of the two-band tight-binding model without strain along paths with fixed $k_z$. The nodal spiral crosses the considered path for two different values of $k_z$.  (f) Tight-binding spectrum of a strained (001) slab for $L=100$, $R=400$ with flat bands. The inset shows the spatial profile of the flat-band state at the red dot, which is again composed of a Landau-level and a bottom-surface state.}
	\label{fig:fig3}
\end{figure*}

In this section, we provide a materials perspective on our proposal.
In particular, we identify CaAgP and rhombohedral graphite NLSMs as possible material platforms for 3D flat bands with negligible spin-orbit interaction.
We note that NLSMs can also be realized in systems with sizable spin-orbit interactions. 
Theoretically proposed candidates are TlTaSe$_2$~\cite{Bian2016} or alloys of SnTe~\cite{Lau2019}.

\subsection{Strained C\lowercase{a}A\lowercase{g}P}

CaAgP crystallizes in an ZrNiAl-type hexagonal structure~\cite{Yamakage:2016_JPSJ}, which is illustrated in Fig.~\ref{fig:fig3}(a).
Standard DFT calculations predict a topological NLSM phase for the 
nodal-line Dirac semimetal CaAgP.
However, standard DFT is known to overestimate the band inversion and, in fact, experimental results for CaAgP show a trivially gapped phase for this compound.
Nonetheless, it has been shown both theoretically and experimentally that strain and As doping can turn the phase from trivial to topological~\cite{Xu:2018_PRB}.
In the following, we focus on the NLSM phase of the CaAgP materials class. 

The low-energy band structure is formed by the $5s$ orbitals of the Ag atoms and the $3p$ orbitals of the light P atoms. 
Therefore, spin-orbit interaction in CaAgP is small.
In the atomic limit, the $p$ orbitals are occupied and the $s$ orbitals are unoccupied.
At the $\Gamma$ and at the A point of the Brillouin zone, there is a band inversion between one $s$ band and one $p_z$ band, while the band orderings at M and K are trivial.
As a consequence, a line node is observed along the paths $\Gamma$-M and $\Gamma$-K, which forms a circle in the $k_x$-$k_y$ plane centred at the $\Gamma$ point [see Fig.~\ref{fig:fig3}(a) and Appendix below].
The band dispersion at the line node is linear along both the radial and the $k_z$ directions.
It should be noted that this line node is generally not protected from spin-orbit interaction but, in this case, the weak spin-orbit coupling induces only a small gap on the order of 10 K~\cite{Yamakage:2016_JPSJ}.

We construct an effective two-orbital tight-binding model (see Appendix below) based on one $3p_z$-orbital Wannier function (WF) centred at one P atom and one $5s$-orbital WF centred within a triangle of Ag atoms as previously 
done for other triangular systems~\cite{Cuono:2019_NJP}.
With an energy cut-off of $6\,\mathrm{meV}$ for the tight-binding parameters,
this allows us to accurately capture the low energy physics with a relatively simple model.
Most importantly, our two-band model correctly reproduces the NLSM phase of the CaAgP materials class [see Fig.~\ref{fig:fig3}(b)].

In the next step, we implement the strain terms by modifying the tight-binding parameters in a way similar to our minimal model above (see Appendix below).
We choose the bending direction to be aligned with the $x$ axis.
We compute the energy spectrum for a (001) slab and find two nearly flat bands, as shown in Fig.~\ref{fig:fig3}(c). Similar to our minimal model, these bands consist of superpositions of bottom-surface states and PLL bulk states. In particular, the center of the PLL bulk states shifts as function of the flat-band momentum, which happens symmetrically around the center of the flat band at $\bar{\Gamma}$. However, due to the broken electron-hole symmetry in this material, the two flat bands acquire a small dispersion. Furthermore, we observe a small energy splitting between the two bands due to chiral-symmetry breaking, a symmetry that was present in the minimal model above due to its simplicity. The corresponding gap is smaller in the graphite case we turn to now.

\subsection{Strained rhombohedral graphite}

Another possible candidate material is rhombohedral graphite.
It consists of light carbon atoms and has, therefore, negligible spin-orbit interaction.
Rhombohedral graphite is composed of ABC-stacked graphene sheets [see Fig.~\ref{fig:fig3}(d)].
Isolated graphene sheets feature Dirac cones at the K and K' points of their 2D Brillouin zone.
Stacking several sheets in the $z$ direction in an ABC-type fashion leads to a $k_z$-dependent shift of the Dirac cones away from the K and K' points thereby forming a nodal line.
In particular, the nodal lines in rhombohedral graphite spiral around the vertical hinges of its hexagonal Brillouin zone [see Fig.~\ref{fig:fig3}(d)].

The low-energy properties of rhombohedral graphite, similar to graphene, can be captured by a two-band model (see Appendix below).
This model reproduces the nodal spirals as can be inferred from comparing the spectrum shown in Fig.~\ref{fig:fig3}(e) to the Brillouin zone sketch in Fig.~\ref{fig:fig3}(d).
Note that the electron-hole symmetry is only weakly broken in this material.

As before, we implement strain terms corresponding to a cylindrical strain profile with the bending direction aligned with the $x$ axis (see Appendix below).
The energy spectrum of a (001) slab of this system features two flat bands at the Fermi level, as illustrated in Fig.~\ref{fig:fig3}(f).
In contrast to CaAgP, their splitting and bandwidth is negligibly small. This is due to the approximate electron-hole symmetry of the material.
The flat bands are again composed of PLL bulk states and drumhead surface states.
However, their behavior and composition as a function of the flat-band momentum $(q_x,q_y)$ differs from our previous observations. Depending on the direction, we observe all possible combinations: bottom-surface and PLL bulk state, PLL bulk and top-surface state, bottom- and top-surface state, and even two PLL bulk states with opposite shift behavior (see Appendix below).
Despite these differences we emphasize that in all cases a 3D flat band appears robustly in the presence of strain.

\section{Discussion and conclusions}
We have proposed a feasible way to create 3D flat bands by applying strain to a nodal-line semimetal. In the process, we have discovered an inherent connection between the arising 3D flat band and the drumhead surface states of the parent nodal-line semimetal. Moreover, we have investigated the effects of interactions on the flat bands, highlighting the competition of superconductivity and magnetism in analogy with twisted bilayer graphene.
By computing the superfluid stiffness, we have confirmed that the flatness of the bands does not impede the phase rigidity of the system.
The flat-band system thus supports supercurrents and true long-range order solely due to the 3D quantum geometry of the Bloch wave functions, which arises because of the variation of the pseudo-magnetic field direction along the nodal line.
Going beyond our general idea, we have applied our theory to the potential candidate materials CaAgP and rhombohedral graphite.
We have shown that both materials give rise to sufficiently flat bands under experimentally accessible strain, therefore representing viable candidate materials for the experimental realization of our ideas.
Our conservative estimates for the critical temperatures are on the order of few Kelvins. However, we emphasize that they depend strongly on the magnitude of the applied strain or the pseudomagnetic field. In particular, if the pseudomagnetic field is controlled with the chemical composition we expect that much stronger pseudofield strengths can be obtained, such that critical temperatures on the order of tens of Kelvins could potentially be achieved. The critical temperatures are proportional to the $\mathbf{k}$-space extent of the flat bands.
Therefore, it is important to note that, even though the portion of the Brillouin zone covered by the flat band is small under the assumptions $Q \ll 1$ and $R \gg L$, the flat bands can generally cover an arbitrary portion of the Brillouin zone. In particular, the first assumption $Q \ll 1$ was made only for the sake of analytical transparency. Moreover, implementing the effective magnetic field by varying the chemical composition so that the radius of the nodal line changes along the $z$-direction, we may also violate the second condition $R \gg L$. In the extreme case $Q \sim 1$ and $R \sim L$ we find numerically that the 3D flat bands can indeed cover most of the Brillouin zone (see Appendix below).

We have considered a specific geometry, realizable by bending the sample, leading to a suitable strain profile.
We have, however, explicitly checked that only the strain tensor component $u_{11}$ matters for the appearance of the 3D flat band (see Appendix below) and, therefore, the only important ingredient for realizing a 3D flat band in nodal-line semimetals is an inhomogeneous tensile strain profile across the sample. Thus, a suitable strain tensor can be realized also in various other ways, such as by utilizing the type of multilayer structure designed in Ref.~\cite{Tang2017} for the creation of a giant linear strain gradient.

We emphasize that if the strain is realized by bending the sample, the different phases and phase transitions can be studied in situ in this system. As discussed above, tuning the strain changes the relative interaction strengths of superconductivity and magnetism thereby allowing to study the transition between these phases.
Moreover, it is known that flat bands are an interesting playground for Anderson localization~\cite{goda2006inverse}. The in-situ tunability of the flat-band degeneracy by strain controls also the effective disorder strength, so that it might be possible to study the Anderson transition in a single sample.

In condensed matter physics, there is the immediate interest for the experimental community to realize this 3D analogue of twisted bilayer graphene in order to study the interplay of quantum geometry, flatness of the dispersion, disorder and intertwining of different types of order. Moreover, our proposal of realizing 3D flat bands by strain engineering nodal-line semimetals opens a new frontier of research beyond condensed matter physics because metamaterials and cold atomic gas systems can be engineered to exhibit dispersion relations of topological semimetals to large precision~\cite{song2019observation, ozawa2019topological}. Controllable interactions in the cold atom systems makes realization of superconductivity and magnetism per our predictions possible in such systems.
Beyond that, there are many related conceptual questions:
how general is the connection between the surface states and the lowest pseudo-Landau levels in the presence of a gauge field.
Can the picture be expanded to other semimetals and probably even to topological insulators, superconductors, weak, and fragile phases?  We also note that Dirac nodal lines in the presence of effective gauge fields can lead to peculiar effective electrodynamics~\cite{nissinen2018}, and it might be interesting to study the connection to surface states also in this context.

\textit{Acknowledgements --} We thank Raquel Queiroz, Leslie Schoop, Roni Ilan, and Adolfo Grushin for useful discussions.
This work was supported by the Foundation for Polish Science through the International Research Agendas program co-financed by the
European Union within the Smart Growth Operational Programme.
We acknowledge the access to the computing facilities of the Interdisciplinary Center of Modeling at the University of Warsaw, Grant No. G73-23 and G75-10.

\textit{Data availability --} The data shown in the figures and the code generating all of the data is available at Ref.~\onlinecite{zenodo}.

\appendix

\begin{widetext}

\section{Strain-induced pseudo-magnetic field}
\label{app:pseudo_field}

In the tight-binding model the strain changes the $x$-bond length, which is implemented by a factor $(1-u_{11})$ in front of $\cos k_{x}$ (see Sec.~\ref{App:StrainImplementation}). Depending on the orbital structure also other modifications may appear, but these non-universal contributions are neglected in our consideration of the two-band NLSM model in this section (For completeness, we consider the  effects of the non-universal orbital-mixing strain terms  in Sec.~\ref{App:orbital_mixing_strain} and demonstrate that they do not significantly influence our results.)
Thus, the Hamiltonian in the presence of strain is assumed to be
\begin{eqnarray}
\tilde{H}(\boldsymbol{k}) &=&
\sigma^{z}[6t-t_{1}-2t (1-u_{11}) \cos k_{x}-2 t \sum_{i=y,z} \cos k_{i}] +2t_{2}\sigma^{x}\sin k_{z} \label{eq:H_0}
\end{eqnarray}

We assume that the radius of the nodal loop $Q$ is small and expand the Hamiltonian near an arbitrary nodal point  $\mathbf{Q}=Q (\cos\theta, \sin\theta) = Q \check{\mathbf{Q}}$, where $\theta$ is the polar angle in momentum space. In this way, we obtain
\begin{eqnarray}
h(\boldsymbol{q}) &=& 2 t \sigma^{z} Q_{x}\left[q_{x}+\frac{u_{11}}{Q}\cos\theta \right] +2 t \sigma^{z} Q_{y}\left[q_{y}+\frac{u_{11}}{Q}\sin \theta \right] +2 t_2 \sigma^{x} q_{z}, \label{continuumH}
\end{eqnarray}
where $\mathbf{q}=q(\cos\theta, \sin \theta) = q \check{\mathbf{Q}}$ describes the deviation of the momentum from the nodal line. Notice that both here and in the following sections  $q$  takes both positive and negative values, \textit{i.e.}, it is not the absolute value of $\mathbf{q}$.
The Hamiltonian $h(\boldsymbol{q})$ is parameterized by three independent parameters: $\theta$, $q$, and $q_{z}$.
The strain term $u_{11}$ can be divided between $q_{x}$ and $q_{y}$ in such a way that the pseudo-magnetic field $\mathbf{B}_5$ is continuous around the nodal loop. To see this, we notice that the strain-induced gauge potential is
\begin{equation}
\boldsymbol{A}_5=-\frac{u_{11} \hbar}{e Q}\left(\cos\theta, \sin\theta,0\right),
\end{equation}
leading to the pseudo-magnetic field given by Eq.~(2) in the main text
\begin{equation}
\mathbf{B}_5=\nabla\times\boldsymbol{A}_5=\frac{\hbar}{eR Q}\left(\sin \theta,-\cos\theta,0\right).
\end{equation}
This can be used to define the pseudo-magnetic length $l_B=\sqrt{\hbar/eB_5}=\sqrt{RQ}$.
The structure of the pseudo-magnetic field in $\mathbf{k}$-space is illustrated in Fig.~\ref{fig:pseudo_field}.

\begin{figure*}
\includegraphics[width=0.85\linewidth]{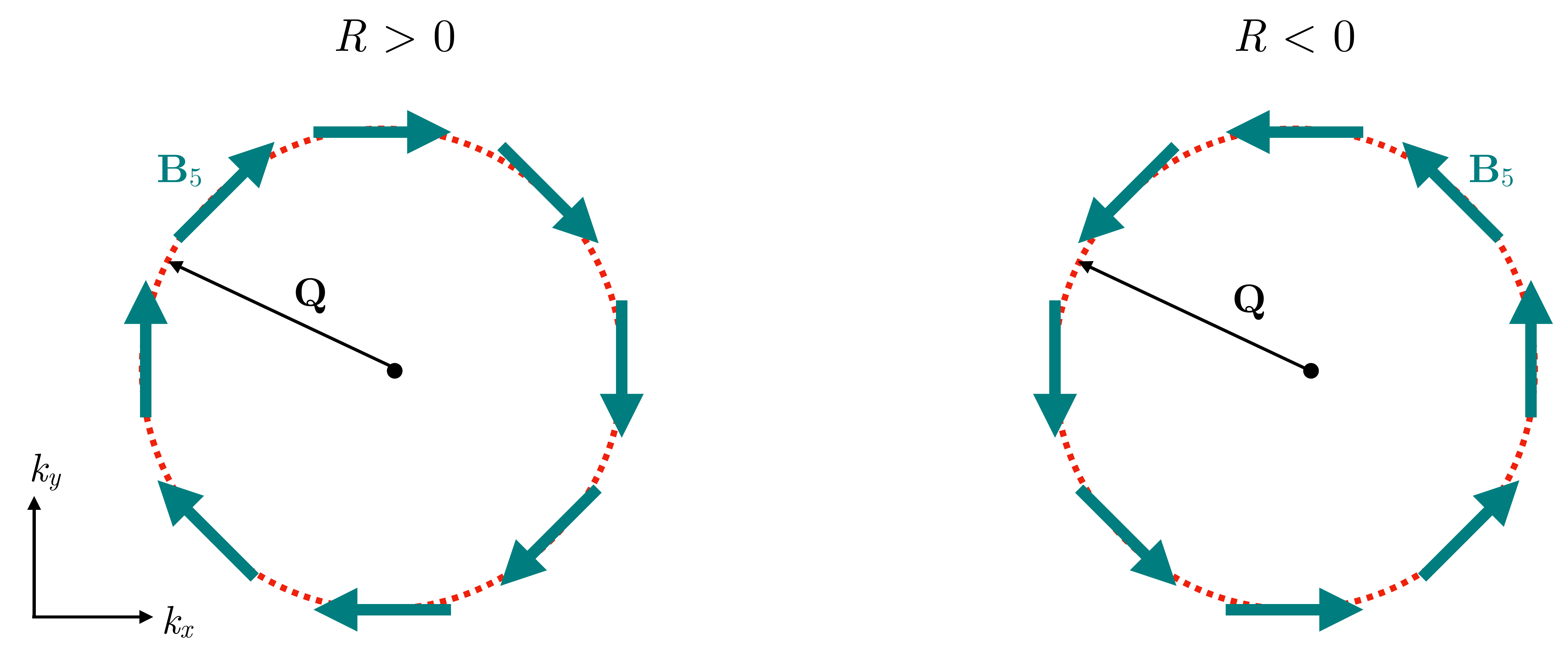}
\caption{Sketch of the momentum-dependent pseudo-magnetic field $\mathbf{B}_5$ along the nodal loop (red dotted line).}
\label{fig:pseudo_field}
\end{figure*}

\section{Bulk wave functions for the lowest pseudo Landau level}

All pseudo-Landau levels of the Hamiltonian (\ref{continuumH}) can be obtained analytically. We skip the explicit derivation here as it is analogous to the derivation of the Landau-level solutions for graphene. We concentrate on the zeroth Landau-level wave functions, which can be written as
\begin{equation}
\Psi_{0}\big(\mathbf{Q}, \mathbf{q} \big)=\frac{1}{\sqrt{2 L_xL_y}} \left(\begin{array}{c}
1\\
i
\end{array}\right)  e^{i(\mathbf{Q}+\mathbf{q})\cdot \mathbf{r}} \left(\frac{1}{\pi l_z^2}\right)^{1/4}e^{-\xi_q^{2}/2}, \label{wavefunction}
\end{equation}
where $\xi_q=(z+z_q)/l_z$, $z_q=q l_B^2$ and the localization length in the $z$-direction is given by
\begin{equation}
l_z=\sqrt{\frac{t_2}{t Q}} l_B. 
\end{equation}
Here, the factor $\sqrt{\frac{t_2}{t Q}}$ describes the elongation of the elliptical semiclassical cyclotron orbit in the $z$-direction.

One can also perform a gauge transformation so that the wave function is localized in the $x-y$ plane and is a plane wave in the $z$-direction. In that case, the localization length is 
\begin{equation}
l_{xy}=\sqrt{\frac{t Q}{t_2}} l_B=\frac{l_B^2}{l_z}.
\end{equation}
Here, the factor $\sqrt{\frac{t Q}{t_2}}$ describes the elongation of the elliptical cyclotron orbit within the $x-y$ plane.
This transformation is in analogy with the gauge transformation that can be performed on Landau-level wavefunctions in two dimensions.

\section{Effective Hamiltonian for the connection of drumhead surface states to lowest pseudo-Landau level}
\label{app:effectiv_Hamiltonian}

To simplify our considerations, we look at solutions along a specific cut through momentum space.
Without loss of generality, we choose $Q_y=0$ for a circular nodal line.
Along this cut, after performing a basis rotation in the space of Pauli matrices of the form $\sigma_z\rightarrow\sigma_x$ and $\sigma_x\rightarrow\sigma_y$, the effective nodal-line Hamiltonian under strain is
\begin{equation}
h(q_x) = 2\sigma_x Q \left( q +\frac{z}{R Q}  \right) - 2i\sigma_y t_2 \partial_z.
\label{eq:hamiltonian}
\end{equation}
By denoting $\Psi_{q}(x,z) = e^{iq x}\,\Phi(z)=e^{iq x}\,(\phi_1,\phi_2)^T$ and assuming $E=0$, we obtain two uncoupled equations
\begin{eqnarray}
\partial_z\phi_1 - \phi_1/l_S +\frac{z}{l_z^2} \phi_1 &= 0,\\
\partial_z\phi_2 + \phi_2/l_S -\frac{z}{l_z^2} \phi_2  &= 0,
\end{eqnarray}
We require that the solutions $\Phi(z)=(\phi_1,\phi_2)^T$ satisfy boundary conditions $\phi_1(z=-L/2)=\phi_2(z=L/2)\approx 0$ up to corrections exponentially small in $1/L$.

\subsection{Drumhead surface states in the unstrained model}

We start by inspecting the solutions of the unstrained limit, \textit{i.e.}, in the limit $l_z \to \infty$.
In this case, the solutions are the drumhead surface states discussed in the main text, which in the new basis can be written as
\begin{eqnarray}
\Phi_\mathrm{top}(z) &= \sqrt{\frac{1}{l_S\sinh(L/l_S)}} 
\left(e^{+z/l_S}, 0\right) \\
\Phi_\mathrm{bot}(z) &= \sqrt{\frac{1}{l_S\sinh(L/l_S)}} 
\left(0, e^{-z/l_S}\right). 
\end{eqnarray}
These solutions are states localized to opposite surfaces (top and bottom) and decay exponentially into the interior of the system.

\subsection{Drumhead surface states and pseudo-Landau levels in the presence of strain}
Once we switch on the strain,  we obtain the following general solutions
\begin{equation}
\Phi(z) = (\phi_1, \phi_2) = 
\left(A\, e^{+z/l_S}\, e^{-\frac{z^2}{2l_z^2}},  
B\, e^{-z/l_S}\, e^{+\frac{z^2}{2 l_z^2}}\right),
\end{equation}
We immediately see that the drumhead-state solutions are recovered in the limit $l_z \rightarrow\infty$.
At finite strain, the drumhead state from the \emph{top surface}, corresponding to $\phi_1$, evolves into a PLL bulk state 
\begin{eqnarray}
\phi_1(z) &= \frac{1}{\sqrt{{\cal N}_1}} \left(\frac{4}{\pi l_z^2}\right)^{1/4} 
e^{-\frac{1}{2l_z^2} \left( z + q l_B^2\right)^2}, \ {\cal N}_1 = \mathrm{Erf}\left(\frac{L+2q l_B^2}{2 l_z}\right) 
+ \mathrm{Erf}\left(\frac{L-2q l_B^2}{2 l_z}\right), 
\end{eqnarray}
where $\mathrm{Erf}(\xi)$ is the Gauss error function. 
Interestingly, this solution satisfies the boundary condition $\phi_1(z=-L/2)\approx 0$ also for $q>0$.
Hence, the radius of the plateau of zero-energy states grows with the system width.

On the other hand, the drumhead state from the \emph{bottom surface}, corresponding to $\phi_2$, does not evolve into a PLL bulk state.
It is modulated with a function $e^{z^2}$ and therefore grows faster than exponentially for $z\rightarrow\pm\infty$.
For sufficiently small strain, we nevertheless expect to recover a state localized to the bottom surface that decays into the interior.
By completing the square and normalizing the wave function, we obtain
\begin{eqnarray}
\phi_2(z) &= \frac{1}{\sqrt{{\cal N}_2}} \left(\frac{4}{\pi l_z^2}\right)^{1/4} 
e^{\frac{1}{2l_z^2} \left( z +  q l_B^2 \right)^2},  \ {\cal N}_2 = \mathrm{Erfi}\left(\frac{L+2 q l_B^2}{2 l_z}\right) 
+ \mathrm{Erfi}\left(\frac{L-2 q l_B^2}{2 l_z}\right), 
\end{eqnarray}
where we have used the imaginary error function $\mathrm{Erfi}(\xi) = -i \mathrm{Erf}(i\xi)$.
As before, this solution is subject to the boundary condition $\phi_2(z=L/2)\approx 0$.
This condition will be satisfied as long as the turning point of $\phi_2$, where its slope changes from negative to positive, is beyond the boundary of the system at $z=L/2$.
This turning point is at $-q l_B^2$.
Hence, $\phi_2$ is only a valid solution for momenta deep inside the nodal circle, not even at $k_x=\pm Q$ ($q_x=0$), which would contradict our numerical findings. This problem can be resolved by defining the approximate analytical solution as
\begin{eqnarray}
\tilde{\phi}_2(z) = 
\begin{cases} 
\frac{1}{\sqrt{\tilde{{\cal N}}_2}}\,e^{-z/l_S}\, e^{+\frac{z^2}{2 l_z^2}} & -L/2\leq z < -q l_B^2 \\
\frac{1}{\sqrt{\tilde{{\cal N}}_2}}\,e^{-\frac{q^2 l_B^4}{2 l_z^2}}  & -q l_B^2 \leq z \leq L/2
\end{cases}
\end{eqnarray}
with
\begin{eqnarray}
\tilde{{\cal N}}_2 = \frac{L +  2  q l_B^2 + \sqrt{\pi} l_z\, \mathrm{Erfi}
		\left( \frac{L - 2 q l_B^2}{2 l_z}\right)}{2e^{q^2 l_B^4/l_z^2}}.
\end{eqnarray}

\subsection{Effective model in the presence of strain}

Using the approximate solutions $\Phi_1 = (\phi_1,0)$ and $\Phi_2 = (0,\phi_2)$, we will now write down an effective Hamiltonian for the flat-band states.
For that, we make use of our insights from numerics: the degenerate states are at zero energy in the interior of the flat band, whereas they split and acquire a finite dispersion
at the flat-band boundary. 
This is due to the Landau-level state shifting towards the bottom surface where the other state is localized.
The two states hybridize due to the surface effects, and we model this by introducing an effective coupling 
\begin{eqnarray}
H_{ij}(q) =
\delta \langle\Phi_i (q)|\sigma_x|\Phi_j (q)\rangle, 
\end{eqnarray}
where the strength of the coupling $\delta$ can be determined by fitting the energies of the effective model $E(q) = \pm H_{12}(q)$ to the numerical results obtained from the full tight-binding Hamiltonian. The coupling $H_{12}(q)$ can be expressed as
\begin{eqnarray}
H_{12}(q) = \frac{\delta}{\sqrt{\pi}} \begin{cases}
\frac{(L - 2q l_B^2)/l_z +\, (2\pi)^{1/2}\, \mathrm{Erf}[(L/2 + q l_B^2)/\sqrt{2 l_z^2}]}
{\sqrt{\left\{\mathrm{Erf}[(L/2 - q l_B^2)/l_z]\, + \,\mathrm{Erf}[(L/2 + q l_B^2)/l_z]\right\}
		\left\{\mathrm{Erfi}[(L/2 - q l_B^2)/l_z] + \pi^{-1/2}\, (L + 2q l_B^2)/l_z \right\}}}, &  -q l_B^2<L/2\\ 
\frac{2L/l_z}
{\sqrt{\left\{\mathrm{Erf}[(L/2 - q l_B^2)/l_z]\, + \,\mathrm{Erf}[(L/2 + q l_B^2)/l_z]\right\}
		\left\{\mathrm{Erfi}[(L/2 - q l_B^2)/l_z] + \mathrm{Erfi}[(L/2 + q l_B^2)/l_z] \right\}}}, &  -q l_B^2 > L/2
\end{cases}
\end{eqnarray}
and the agreement with the numerical results is excellent as shown in the main text.

\section{Mean field theory of flat-band magnetism \label{App:FM}}

We start by discussing the PLL ferromagnetism in the bulk. We point out that this theory describes several different types of magnetic order parameters because their projections to PLL bulk wave functions are the same. However, the surface effects distinguish some of these magnetic order parameters from each other, and in the end of the section we identify the magnetic order parameters favoured by the surface effects.

The Coulomb interactions projected to the zeroth PLL wave functions [Eq.~(\ref{wavefunction})] are described by the Hamiltonian 
\begin{equation}
\hat{H}_{I}=\frac{1}{2}\sum_{\sigma,\sigma'}\sum_{\mathbf{Q}, \mathbf{Q}'} \sum_{\mathbf{q}, \mathbf{q}'} \sum_{\mathbf{K}, \mathbf{k}} V_P(\mathbf{q}, \mathbf{q}', \mathbf{K}, \mathbf{k})\hat{c}_{\mathbf{Q}, \mathbf{q}, \sigma}^\dagger\hat{c}_{\mathbf{Q}', \mathbf{q}', \sigma'}^\dagger\hat{c}_{\mathbf{Q}'+\mathbf{K}, \mathbf{q}'+\mathbf{k}, \sigma'} \hat{c}_{\mathbf{Q}-\mathbf{K}, \mathbf{q}-\mathbf{k}, \sigma},
\label{eq:coulomb_projected}
\end{equation}
\begin{equation}
V_P(\mathbf{q}, \mathbf{q}', \mathbf{K}, \mathbf{k})=  \int d^2 r \int dz \int dz' V_C(\mathbf{r}, z-z') \frac{e^{i (\mathbf{K+k}) \cdot \mathbf{r}}}{L_x L_y} \phi_0\bigg(\frac{z+z_{\mathbf{q}}}{l_z}\bigg) \phi_0\bigg(\frac{z'+z_{\mathbf{q}'}}{l_z}\bigg) \phi_0\bigg(\frac{z'+z_{\mathbf{q}'+\mathbf{k}}}{l_z}\bigg) \phi_0\bigg(\frac{z+z_{\mathbf{q}-\mathbf{k}}}{l_z}\bigg), \nonumber
\end{equation}
where $\sigma$ describes the spin of the electron,
\begin{equation}
\phi_0(\xi)=\frac{1}{\pi^{1/4} \sqrt{l_z}} e^{-\xi^2/2},
\label{eq:phi0_definition}
\end{equation}
and
\begin{equation}
V_C(\mathbf{r}, z)=\frac{e^2}{4 \pi \epsilon \epsilon_0 \sqrt{r^2+z^2}}.
\end{equation}
Here the summations are subject to the restrictions: 
\begin{itemize}
	\item $\mathbf{Q}$, $\mathbf{Q}'$, $\mathbf{Q}'+\mathbf{K}$ and $\mathbf{Q}-\mathbf{K}$ are on the nodal line,
	\item $\mathbf{q}$ and $\mathbf{q}'$ are perpendicular to the nodal line at the corresponding $\mathbf{Q}$ and $\mathbf{Q}'$, respectively,
	\item $\mathbf{q}'+\mathbf{k}$ and $\mathbf{q}-\mathbf{k}$ are perpendicular to the nodal line at the points described by $\mathbf{Q}'+\mathbf{K}$ and $\mathbf{Q}-\mathbf{K}$, respectively.
\end{itemize}

The Hartree-Fock approximation for $\hat{H}_I$, assuming spatially homogeneous ferromagnetism, can then be implemented by assuming that 
\begin{equation}
\langle c_{\mathbf{Q}, \mathbf{q}, \sigma}^\dag c_{\mathbf{Q'}, \mathbf{q'}, \sigma'} \rangle \ne 0 \iff \mathbf{Q}=\mathbf{Q'} \textrm{, } \mathbf{q}=\mathbf{q'} \textrm{ and } \sigma=\sigma'.
\end{equation}
The direction of the magnetization can be chosen arbitrarily due to the $SU(2)$-symmetry, and here we have chosen it to be along the $z$-direction. This way we obtain
\begin{eqnarray}
\hat{H}_{I}&=&\sum_{\sigma,\sigma'}\sum_{\mathbf{Q}, \mathbf{Q}'} \sum_{\mathbf{q}, \mathbf{q}'} V_P(\mathbf{q}, \mathbf{q}', \mathbf{0}, \mathbf{0})  \langle\hat{c}_{\mathbf{Q}, \mathbf{q}, \sigma}^\dagger \hat{c}_{\mathbf{Q}, \mathbf{q}, \sigma}\rangle \hat{c}_{\mathbf{Q}', \mathbf{q}', \sigma'}^\dagger\hat{c}_{\mathbf{Q}', \mathbf{q}', \sigma'}  \nonumber \\
&&\hspace{-0.25cm}-\sum_{\sigma}\sum_{\mathbf{Q}, \mathbf{Q}'} \sum_{\mathbf{q}, \mathbf{q}'} V_P(\mathbf{q}, \mathbf{q}', \mathbf{Q-Q'}, \mathbf{q-q'}) \langle \hat{c}_{\mathbf{Q}, \mathbf{q}, \sigma}^\dagger \hat{c}_{\mathbf{Q}, \mathbf{q}, \sigma} \rangle \hat{c}_{\mathbf{Q}', \mathbf{q}', \sigma}^\dagger \hat{c}_{\mathbf{Q}', \mathbf{q}', \sigma} +\textrm{Const.} \nonumber \\
\end{eqnarray}

We now additionally assume that the density is a constant independent of the position
\begin{equation}
\sum_\sigma \langle c_{\mathbf{Q}, \mathbf{q}, \sigma}^\dag c_{\mathbf{Q}, \mathbf{q}, \sigma}  \rangle = C, \label{self-cons-mu}
\end{equation}
where $0 \leq C \leq 2$ is the filling factor of the flat bands.

\begin{figure*}
\includegraphics[width=0.85\linewidth]{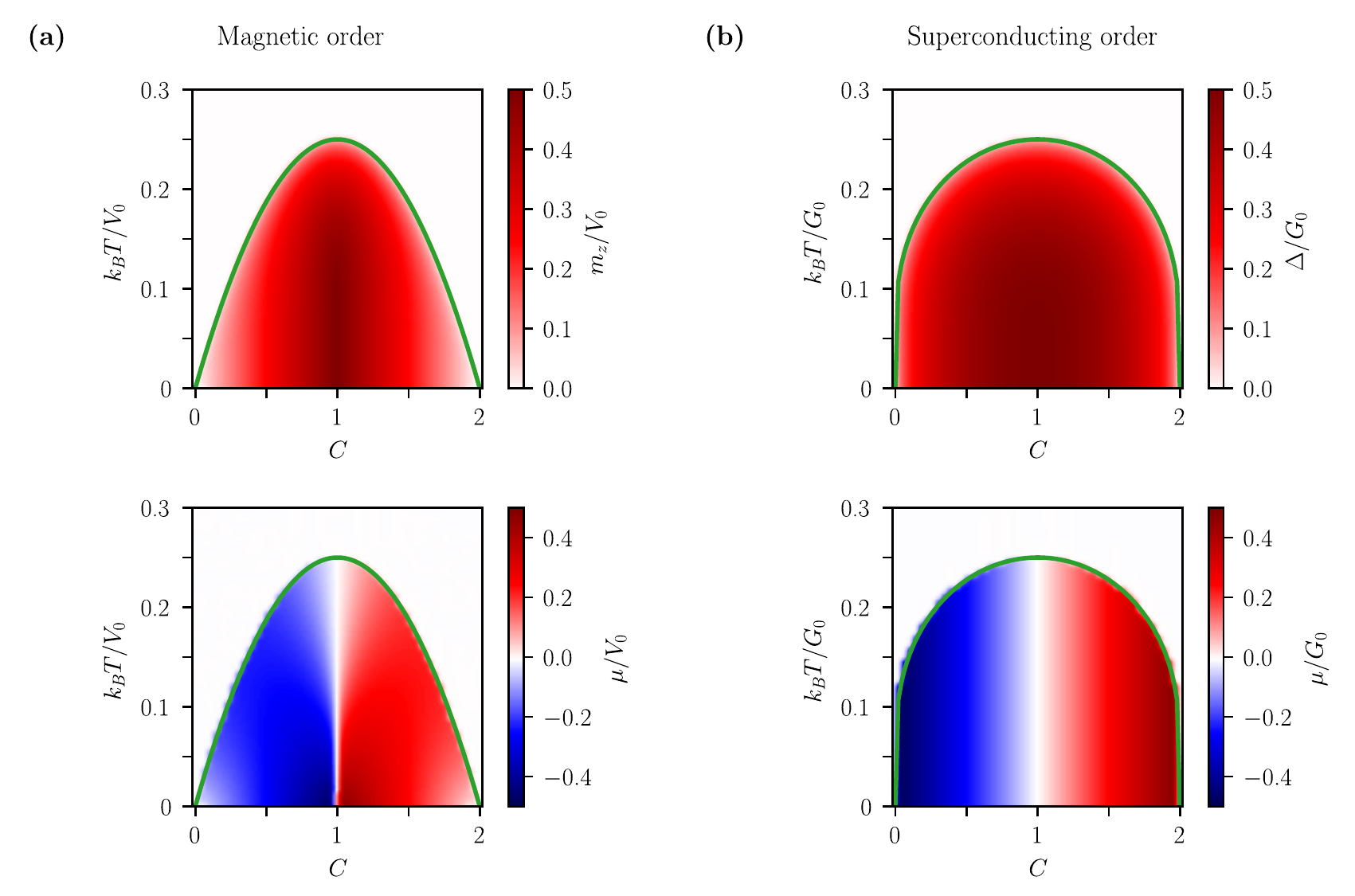}
\caption{Phase diagrams based on the mean-field equations for (a) magnetic order and for (b) superconducting order. $V_0$ and $G_0$ are the interaction strengths, $m_z$ is the magnetization, $\Delta$ is the superconducting gap, $\mu$ is the chemical potential, and $C$ is the density. The green lines correspond to the analytical formulas of the critical temperatures $T_c$.}
\label{fig:phase_diagrams}
\end{figure*}

Then the mean field Hamiltonian simplifies to a form
\begin{equation}
\hat{H}_{\textrm{mf}}=\sum_{\mathbf{Q}, \mathbf{q}} C_{\mathbf{Q}, \mathbf{q}}^\dag H_{\textrm{mf}} C_{\mathbf{Q}, \mathbf{q}},
\end{equation}
where $C_{\mathbf{Q}, \mathbf{q}}^\dag=(c_{\mathbf{Q}, \mathbf{q}, \uparrow}^\dag, c_{\mathbf{Q}, \mathbf{q}, \downarrow}^\dag)$, 
\begin{equation}
H_{\textrm{mf}}=m_z \sigma_z-\mu \sigma_0 
\end{equation}
and
\begin{equation}
m_z=-\frac{1}{2}\sum_{\mathbf{Q}'} \sum_{\mathbf{q}'} V_P(\mathbf{q}, \mathbf{q}', \mathbf{Q-Q'}, \mathbf{q-q'}) \bigg[ \langle \hat{c}_{\mathbf{Q}', \mathbf{q}', \uparrow}^\dagger \hat{c}_{\mathbf{Q}', \mathbf{q}', \uparrow} \rangle - \langle \hat{c}_{\mathbf{Q}', \mathbf{q}', \downarrow}^\dagger \hat{c}_{\mathbf{Q}', \mathbf{q}', \downarrow} \rangle \bigg]. \label{self-cons-m}
\end{equation}
The magnetization $m_z$ is independent of $\mathbf{Q}$ and $\mathbf{q}$ due to the spatial homogeneity, and thus $m_z$ and the chemical potential $\mu$ should be solved self-consistently using Eqs.~(\ref{self-cons-mu}) and (\ref{self-cons-m}).
By straightforward calculation we obtain the mean-field equations 
\begin{eqnarray}
m_z &=&  \frac{V_0}{2}\, \frac{\sinh\beta m_z}{\cosh \beta\mu + \cosh \beta m_z}, 
\label{eq:ferro_mf_1}\\
C &=&  1 + \frac{\sinh\beta\mu}{\cosh \beta\mu + \cosh \beta m_z},
\label{eq:ferro_mf_2}
\end{eqnarray}
where $\beta = 1/k_B T$ and we have defined
$V_0 = \sum_{\mathbf{Q}'} \sum_{\mathbf{q}'} V_P(\mathbf{q}, \mathbf{q}', \mathbf{Q-Q'}, \mathbf{q-q'})$.
We solve these equations numerically by reformulating them in terms of a minimization problem, for which we then compute the minima using a stochastic algorithm based on Basin-hopping.
The results are presented in Fig.~\ref{fig:phase_diagrams}(a).
Furthermore, from the equations above it is easy to see that the critical temperature for magnetism depends on the filling factor as
\begin{equation}
k_B T_{c, m} = \frac{V_0}{4} C (2-C). \label{Tc-FM}
\end{equation}

At zero temperature we obtain 
\begin{equation}
m_z(T=0)=\frac{V_0}{2} (1-|C-1|).
\end{equation}

\begin{figure*}
\includegraphics[width=0.85\linewidth]{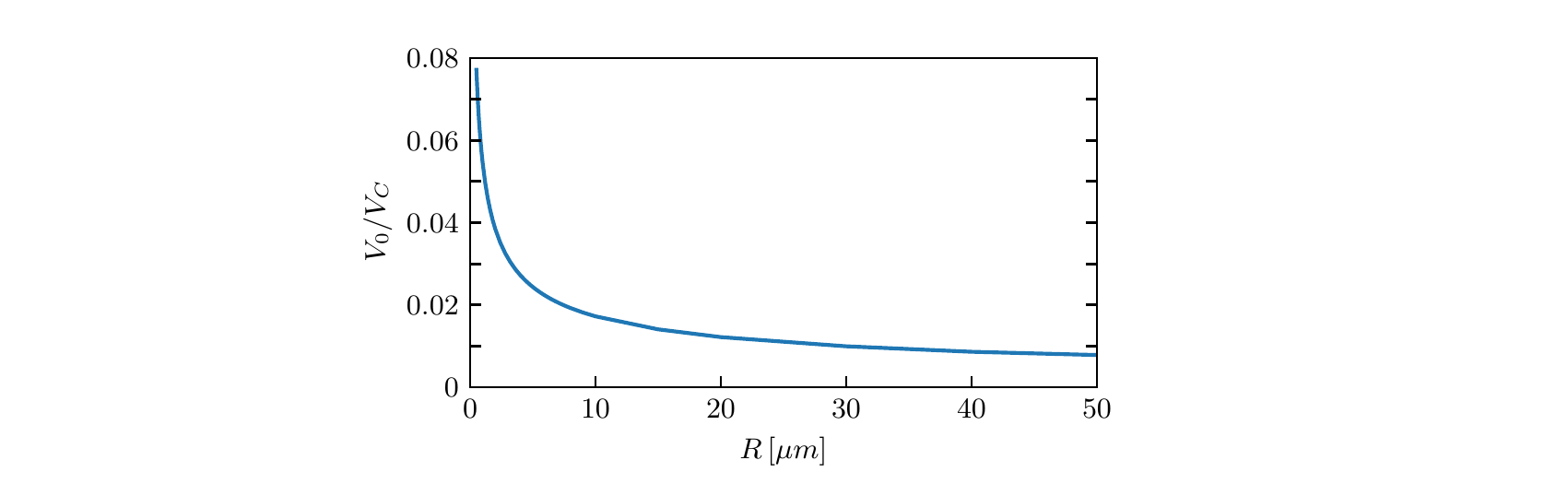}
\caption{Interaction strength $V_0$ as a function of the curvature radius $R$ for the minimal model with $t=0.21\,\mathrm{eV}$, $t_1=0.25t$, $t_2=0.8t$, and the lattice constant set to $d=0.97\,\mathrm{nm}$.}
\label{fig:magnetic_V0_scaling}
\end{figure*}

The interaction strength depends on the parameters of the model as
\begin{equation}
V_0 =  \frac{Q V_C d}{\sqrt{2}}
\int dq'
\int rdr\: \phi_0^2\bigg(\frac{q' l_B^2}{\sqrt{2}l_z}\bigg) 
\exp\!\bigg(\frac{r^2}{4l_z^2}\bigg) K_0\!\bigg(\frac{r^2}{4l_z^2}\bigg)
J_0[(Q + q')r]\,
J_0(Qr),
\end{equation}
where $V_C=e^2/(4\pi\epsilon\epsilon_0 d)$ with the lattice constant $d$.
Figure~\ref{fig:magnetic_V0_scaling} shows the numerically calculated $V_0$ as a function of the curvature radius $R$. We have used model parameters estimated from a fit to the nodal line in CaAgP (see caption of Fig.~\ref{fig:magnetic_V0_scaling}), with the parameter $t_1$ tuned from $0.85t$ to $0.25t$ to obtain a smaller nodal circle in agreement with our initial assumptions of sufficiently small $Q$.  We find that $V_0$ scales like $1/\sqrt{R}$.  
By using $R=0.5\,\mathrm{\mu m}$, we therefore estimate that $V_0=114.5\,\mathrm{meV}\,/\epsilon$. We further estimate that typical values of the dielectric constant $\epsilon$ for these materials are on the order of 100. Hence, critical temperatures can be on the order of $T_{c,m}=3\,\mathrm{K}$.

Notice that we have assumed a larger dielectric constant than typically observed in bulk semimetals~\cite{schilling2017flat}. The reason is that the enhanced density of states in our system is expected to lead to larger screening effects. We also point out that the relevant length scales in our problem are similar to those in twisted bilayer graphene and our estimate for the critical temperature agrees with the experimentally observed critical temperatures in that system.
Thus, our estimate can be considered conservative guided by the current knowledge about its 2D analogue, twisted bilayer graphene, but it might also be possible to observe larger critical temperatures in nodal-line semimetals due to larger stability to fluctuations in 3D and larger variability of parameters with strain.

We point out that our calculation is compatible with various types of spatially uniform magnetic orders because they can lead to the same projected order parameter within the low-energy theory. For this purpose we now consider the order parameters of the form $M = m_j \sigma_i s_j$, where  Pauli matrices $\sigma_i$ ($i=0,x,y,z$) and $s_j$ ($j=x,y,z$) correspond to the orbital and spin degrees of freedom, respectively.
Due to the $SU(2)$ spin symmetry of the nonmagnetic phase, we can restrict our considerations to the case $j=z$.

\begin{figure*}
\includegraphics[width=0.85\linewidth]{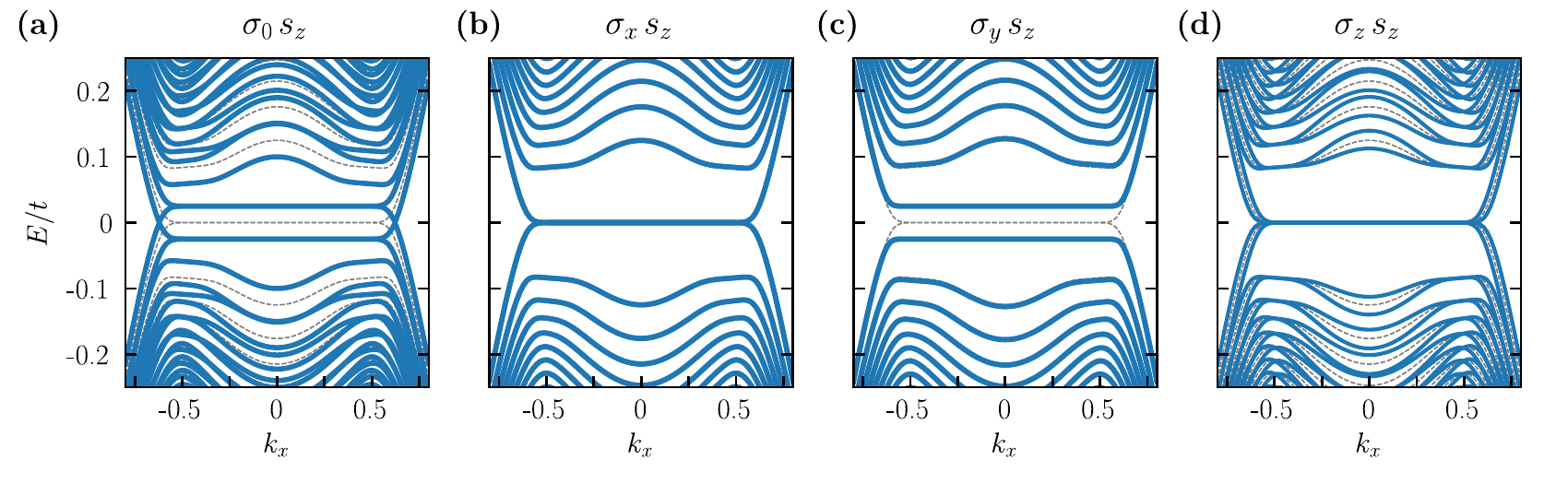}
\caption{Magnetic spectra of the minimal model in a (001) slab geometry with parameters $t_1=0.25t$, $t_2=0.8t$, $R=800$, $L=200$, and $m_z=0.025t$. The considered structure $\sigma_i s_j$ of the order parameter is indicated. The corresponding spectra for $m_z=0$ are plotted with dashed lines. Only the spectrum in (c) is fully gapped.}
\label{fig:magnetic_spectra}
\end{figure*}

The zeroth PLL wavefunctions are eigenstates of $\sigma_y$, and therefore for order parameters with $\sigma_x s_z$ and $\sigma_z s_z$ the bulk PLL states stay at zero energy such that the system remains gapless. Thus, these order parameters are not energetically favored.
On the other hand, order parameters $\sigma_0 s_z$ and $\sigma_y s_z$  lead to the same projected order parameter within the PLL states and open a gap in the bulk. Thus, both of these order parameters are compatible with the calculation given above and good candidates for the ground state. However, there is a further distinction between these order parameters when the surface effects are taken into account as shown in Fig.~\ref{fig:magnetic_spectra}. The order parameter $\sigma_y s_z$ gives rise to a full gap both in the bulk and at the surface, but the order parameter $\sigma_0 s_z$ just shifts the two spin blocks of the Hamiltonian oppositely in energy so that the two shifted sets of bands cross at the edge of the flat bands. Therefore, we expect that the order parameter $\sigma_y s_z$  will be energetically favored.

To shed light on the structure of this order parameter, we perform a change of basis through $\tilde{M} = U^\dagger M U$ with 
\begin{equation}
U = \frac{1}{\sqrt{2}}
\begin{pmatrix}
1 & 1\\
i & -i\\
\end{pmatrix} \otimes s_0.
\end{equation}
In this basis the order parameter $\tilde{M}=m_z\, \tilde{\sigma}_z s_z $ is diagonal, and the magnetic order is staggered with respect to PLL bulk states and drumhead states at the bottom surface.
In particular, if we restrict ourselves to the subspace of PLL bulk states, this corresponds to a ferromagnetic order parameter.
Therefore, the calculations for the magnetic phase of the PLL bulk states, presented at the beginning of this section, is fully compatible also with respect to this order parameter.

\section{Mean field theory for flat-band superconductivity \label{App:SC}}

When considering an attractive interaction between electrons within the reduced BCS Hamiltonian approach, the pairing Hamiltonian takes the form
\begin{equation}
H_{\rm pairing}=\sum_{\mathbf{Q}, \mathbf{Q}', \mathbf{q}, \mathbf{q}'} G_P(q,q') c^\dag_{\mathbf{Q} \mathbf{q} \uparrow} c^\dag_{-\mathbf{Q} -\mathbf{q} \downarrow} c_{\mathbf{Q}' \mathbf{q}' \uparrow} c_{-\mathbf{Q}' -\mathbf{q}' \downarrow},
\end{equation}
where the projected two-particle interaction potential, corresponding to an effective on-site attraction $V(\mathbf{r}_1, \mathbf{r}_2)=-g \delta(\mathbf{r}_2-\mathbf{r}_1)$, is
\begin{equation}
G_P(q,q')=\frac{g}{L_x L_y} \int dz \ \phi_0^2\bigg(\frac{z+z_q}{l_z} \bigg) \phi_0^2\bigg(\frac{z+z_{q'}}{l_z} \bigg).
\end{equation}
Here $g>0$ corresponds to the attractive interaction. Therefore, the gap equation can be written as
\begin{equation}
\Delta(\mathbf{Q}, \mathbf{q})=\sum_{\mathbf{Q}' \mathbf{q}'}  G_P(q,q') \langle c_{\mathbf{Q}' \mathbf{q}' \uparrow} c_{-\mathbf{Q}' -\mathbf{q}' \downarrow} \rangle.
\end{equation}
This equation should be solved self-consistently in the presence of the density constraint
\begin{equation}
C=2 \sum_{\mathbf{Q} \mathbf{q}}   \langle c^\dag_{\mathbf{Q} \mathbf{q} \uparrow} c_{\mathbf{Q} \mathbf{q} \uparrow} \rangle.
\end{equation}

Assuming a homogeneous order parameter $\Delta(\mathbf{Q}, \mathbf{q})=\Delta$, we obtain
\begin{equation}
\Delta=G_{0} \frac{\Delta}{2 \sqrt{\mu^2+\Delta^2}} \tanh(\beta \sqrt{\Delta^2+\mu^2}/2). \label{mean-field-delta}
\end{equation}
and
\begin{equation}
C=1+\frac{\mu}{\sqrt{\mu^2+\Delta^2}} \tanh(\beta \sqrt{\Delta^2+\mu^2}/2),
\end{equation}
where we have defined 
\begin{equation}
G_{0}=\sum_{\mathbf{Q}' \mathbf{q}'}  G_P(0,q')= \frac{gQ}{2\pi l_B^2}. 
\end{equation}

From these equations one obtains a general solution for $\mu$ (valid at all temperatures) given by
\begin{equation}
\mu=\frac{G_0}{2} (C-1)
\end{equation}
and then $\Delta_0$ can be solved from Eq.~(\ref{mean-field-delta}) as a function of temperature.
At zero temperature we get
\begin{equation}
\Delta(T=0)=\frac{G_0}{2} \sqrt{C(2-C)}
\end{equation}
and the critical temperature is given by
\begin{equation}
k_BT_{c, sc}=\frac{G_0}{4} \frac{C-1}{{\rm arctanh}(C-1)}. \label{Tc-SC}
\end{equation}
Moreover, we have also solved the mean-field equations numerically [see Fig.~\ref{fig:phase_diagrams}(b)].

Similarly as in the case of magnetism, the largest order parameter $\Delta(T=0)=G_0/2$ and critical temperature $k_BT_{c, sc}=G_0/4$ for superconductivity are obtained at half filling, but the important qualitative difference is that when the system is doped away from half filling the critical temperature for superconductivity decreases more slowly than the critical temperature for magnetism. For typical model parameters (see caption of Fig.~\ref{fig:magnetic_V0_scaling}),  $R=0.5\,\mathrm{\mu m}$, and $g=1.0\,\mathrm{eV\,nm}^3$, the effective interaction strength is $G_0=0.3\,\mathrm{meV}$ indicating that the critical temperature can be on the order of $T_{c, sc}=1\,\mathrm{K}$.

\section{Out-of-plane superfluid stiffness \label{App:out-of-plane-stiffness}}

In this section, we calculate the superfluid stiffness in the $z$-direction by studying the energy cost of creating a phase-gradient $\Delta(z)=\Delta_0 \exp(i k z)$ similarly as in Refs.~\onlinecite{Moon1995, Pikulin2016}.
In our formalism, we can conveniently study this by assuming that $f(z)=f(-q l_B^2)=\langle c_{\mathbf{Q} \mathbf{q} \uparrow} c_{-\mathbf{Q} -\mathbf{q} \downarrow} \rangle=f_0 \exp(-i k q l_B^2)$. For simplicity, we assume $T \to 0$, $C \to 1$ and $k \to 0$, so that $f_0=1/2$ and the energy is
\begin{eqnarray}
\langle H_{\rm pairing} \rangle&=&\sum_{\mathbf{Q}, \mathbf{Q}', \mathbf{q}, \mathbf{q}'} G_P(q,q') \langle c^\dag_{\mathbf{Q} \mathbf{q} \uparrow} c^\dag_{-\mathbf{Q} -\mathbf{q} \downarrow} \rangle  \langle c_{\mathbf{Q}' \mathbf{q}' \uparrow} c_{-\mathbf{Q}' -\mathbf{q}' \downarrow} \rangle  
=-\frac{1}{4} \sum_{\mathbf{Q}, \mathbf{Q}', \mathbf{q}, \mathbf{q}'} G_P(q,q') \exp(i k (q-q') l_B^2) \nonumber \\
&\approx& -\frac{1}{2} \sum_{\mathbf{Q}, \mathbf{q}} \Delta   + \frac{1}{8} \sum_{\mathbf{Q}, \mathbf{Q}', \mathbf{q}, \mathbf{q}'} G_P(q,q') k^2 (q-q')^2 l_B^4= E_c  + \frac{1}{2} V D_{s,z} k^2 , 
\end{eqnarray}
where $E_c= -\frac{1}{2} \sum_{\mathbf{Q}, \mathbf{q}} \Delta=-V n_0 \Delta/2$ is the condensation energy, $V$ is the volume, $D_{s,z}$ is the superfluid weight in the $z$-direction, and $n_0=Q/(2 \pi l_B^2)$ is the density of particles within the flat band.
In this way, we identify
\begin{eqnarray}
D_{s,z} &=& \frac{n_0 l_B^2}{4} \sum_{\mathbf{Q}, \mathbf{q}} G_P(0,q) q^2 l_B^2 =\frac{n_0 l_z^2}{2}  \Delta. \label{out-of-stiffness}
\end{eqnarray}

\section{In-plane superfluid stiffness \label{App:in-plane-stiffness}}

The in-plane superfluid stiffness $D_s$ also relates the supercurrent $\mathbf{j}$ in a superconductor to a gauge potential $\mathbf{A}$ (in Coulomb gauge) by
\begin{equation}
j_i = -\frac{4 e^2}{\hbar^2}\sum_j [D_s]_{ij} A_j.
\end{equation}
Here we have introduced a prefactor $4e^2/\hbar^2$ so that we have a common convention with Sec.~\ref{App:out-of-plane-stiffness}, where the stiffness was determined from the energy cost. Although the stiffness is generally a tensor, in the case studied here the off-diagonal components vanish, so that we are only interested in the stiffness in different directions.

In general the stiffness is composed of a conventional and a geometrical contribution
\begin{equation}
D_s = D_{s,\mathrm{conv}} + D_{s,\mathrm{geom}},
\end{equation}
but due to the dispersionless flat band the conventional contribution vanishes
\begin{equation}
D_{s,\mathrm{conv}} = 0.
\end{equation}
The geometric contribution at half-filling $C=1$ can be calculated as~\cite{Peotta2015, Liang2017, Xiang2019, Xie2020, Julku2020, hu2020quantummetricenabled}
\begin{equation}
[D_s]_{ij} =  \Delta  \frac{1}{L}
\int\frac{d^2 k}{(2\pi)^2}\: g_{ij}(\mathbf{k}),
\label{eq:def_superfluid_weight}
\end{equation}
where $g_{ij}(\mathbf{k})$ is the Fubini-Study metric of the spin-up  Bloch wave functions $u(\mathbf{k})$.
It is related to the quantum geometric tensor
\begin{equation}
G_{ij} = \partial_{k_i}u^\dagger(\mathbf{k}) [1 - u(\mathbf{k})u^\dagger(\mathbf{k})] \partial_{k_j}u(\mathbf{k}),
\end{equation}
through the relation
\begin{equation}
g_{ij} = \mathrm{Re}(G_{ij}).
\label{eq:gij_Gij}
\end{equation}
The Berry curvature of the system is also related to this quantity by
\begin{equation}
F_{xy} = 2\,\mathrm{Im}(G_{xy}),
\end{equation}
and it can be shown that
\begin{equation}
\mathrm{tr}\,g = g_{xx} + g_{yy} \geq |F_{xy}|.
\end{equation}
Hence, non-vanishing Berry curvature provides a lower bound for the geometric contribution to the superfluid weight.

In the following, we analytically calculate the geometric contribution to the superfluid weight for our model Hamiltonian assuming that the dominant contribution comes from the zeroth PLL bulk states. The corresponding Bloch wave function is 
\begin{equation}
\langle z|u_\mathbf{k}\rangle  = u(\mathbf{k},z) = 
\frac{1}{\sqrt{2}}
\left(\begin{array}{c}
1\\
i
\end{array}\right)
\left(\frac{1}{\pi l_z^2}\right)^{1/4}e^{-\xi_\mathbf{k}^{2}(z)/2},
\end{equation}
with 
\begin{equation}
\xi_\mathbf{k}(z) = \frac{z+z_\mathbf{k}}{l_0} = \frac{z+(k-Q)l_B^2}{l_z},
\end{equation}
where we have now used polar coordinates for $\mathbf{k} = (k\cos{\alpha}, k\sin{\alpha})$.

The Fubini-Study metric $g_{ij}$ of the flat-band states can be expressed as 
\begin{equation}
g_{ij} = \frac{1}{2} \Big( \int dz \left[ \partial_{k_i} u^\dagger(\mathbf{k},z)\, \partial_{k_j} u(\mathbf{k},z) +  \partial_{k_j} u^\dagger(\mathbf{k},z)\, \partial_{k_i} u(\mathbf{k},z) \right] \Big) + \int dz \int dz'\, u^\dagger(\mathbf{k},z)\, \partial_{k_i} u(\mathbf{k},z)\,
u^\dagger(\mathbf{k},z')\, \partial_{k_j} u(\mathbf{k},z').
\end{equation}
The components of this quantity contain the following terms,
\begin{eqnarray}
\partial_{k_x} u(\mathbf{k},z) =
\cos{\alpha} \,\partial_{k} u(k,z) = -u(k,z)\, \frac{l_B^2}{l_z^2}\,\cos{\alpha}\, (z + z_k)\\
\partial_{k_y} u(\mathbf{k},z) =
\sin{\alpha} \,\partial_{k} u(k,z) = -u(k,z)\, \frac{l_B^2}{l_z^2}\,\sin{\alpha}\, (z + z_k)\\
\end{eqnarray}
Using $u^\dagger u = \exp\left[-(z+z_k)^2/l_z^2\right]/\sqrt{\pi}l_z$, we further get
\begin{eqnarray}
\partial_{k_x} u^\dagger\partial_{k_x} u &=& 
\frac{1}{\sqrt{\pi}} \frac{l_B^4}{l_z^5}\cos^2\alpha\, e^{-(z+z_k)^2/l_z^2}\,(z+z_k)^2\\
\partial_{k_y} u^\dagger\partial_{k_y} u &=& 
\frac{1}{\sqrt{\pi}} \frac{l_B^4}{l_z^5}\sin^2\alpha\, e^{-(z+z_k)^2/l_z^2}\,(z+z_k)^2\\
\partial_{k_x} u^\dagger\partial_{k_y} u &=& 
\frac{1}{2\sqrt{\pi}} \frac{l_B^4}{l_z^5}\sin{2\alpha}\, e^{-(z+z_k)^2/l_z^2}\,(z+z_k)^2
= \partial_{k_y} u^\dagger\partial_{k_x} u\\
u^\dagger\partial_{k_x} u &=&
-\frac{1}{\sqrt{\pi}} \frac{l_B^2}{l_z^3}\cos{\alpha}\, e^{-(z+z_k)^2/l_z^2}\,(z+z_k)\\
u^\dagger\partial_{k_y} u &=&
-\frac{1}{\sqrt{\pi}} \frac{l_B^2}{l_z^3}\sin{\alpha}\, e^{-(z+z_k)^2/l_z^2}\,(z+z_k).
\end{eqnarray}
From the last two equations we see that the second term in $g_{ij}$ is zero as it contains only symmetric integrals over odd functions in $z$ and $z'$.
Hence, the components of the Fubini-Study metric $g_{ij}$ simplify to
\begin{eqnarray}
g_{xx} 
&=& \int dz\, \partial_{k_x} u^\dagger\, \partial_{k_x} u
= \frac{1}{2} \frac{l_B^4}{l_z^2}\cos^2{\alpha}\\
g_{yy} 
&=& \int dz\, \partial_{k_y} u^\dagger\, \partial_{k_y} u
= \frac{1}{2} \frac{l_B^4}{l_z^2}\sin^2{\alpha}\\
g_{xy} 
&=& \int dz\, \partial_{k_x} u^\dagger\, \partial_{k_y} u
= \frac{1}{4} \frac{l_B^4}{l_z^2}\sin{2\alpha} = g_{yx}.
\end{eqnarray}

To evaluate the $k$-space integrals of $g_{ij}=g_{ij}(\alpha)$ we consider a system extending from $z=-L/2$ to $+L/2$. The PLL states at $k=Q$ are centered at $z=0$ and reach the top (bottom) surface at $k_\mathrm{top} = Q-L/2l_B^2$ ($k_\mathrm{bot} = Q+L/2l_B^2$).
This defines our domain of integration and we obtain
\begin{eqnarray}
\int_{\mathrm{LLs}}  d^2 k\: g_{ij} &=& \int_{k_\mathrm{top}}^{k_\mathrm{bot}} kdk \int_0^{2\pi}d\alpha\:g_{ij}(\alpha)
= \frac{1}{2}\left( k_\mathrm{bot}^2 - k_\mathrm{top}^2 \right) \int_0^{2\pi}d\alpha\:g_{ij}(\alpha)
= \delta_{ij} \frac{\pi}{2}\frac{l_B^2}{l_z^2}Q \, L.
\label{eq:in-stiffness_integral}
\end{eqnarray}
Thus, $D_{ij}$ is isotropic and has vanishing off-diagonal components. The in-plane superfluid stiffness is given by
\begin{equation}
D_{s,\parallel} =  \Delta  \frac{1}{8 \pi} \frac{l_B^2}{l_z^2}Q=  \Delta  \frac{n_0}{4} \frac{l_B^4}{l_z^2}=  \Delta  \frac{n_0}{4} l_{xy}^2. \label{in-stiffness}
\end{equation}
The expressions for the out-of-plane [Eq.~(\ref{out-of-stiffness})] and in-plane [Eq.~(\ref{in-stiffness})] stiffness are related to each other by replacement of $l_z^2$ with $l_{xy}^2/2$ which arises because of the different elongations of the semiclassical cyclotron orbits in the $z$-direction and within the $(x,y)$-plane and due to the angular average within the $(x,y)$-plane.
Similar results for the superfluid stiffness are expected also beyond mean-field approximation~\cite{peri2020fragile}.

\begin{figure*}
\includegraphics[width=0.85\linewidth]{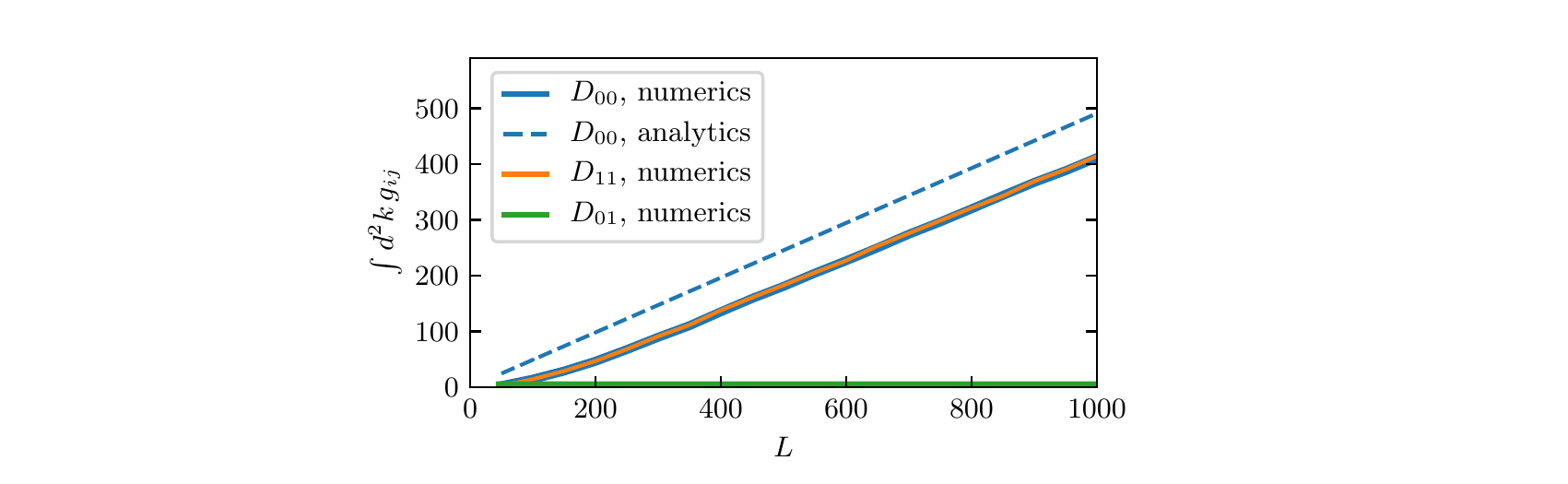}
\caption{Scaling of the in-plane superfluid stiffness for the minimal model with parameters $t=0.21\,\mathrm{eV}$, $t_1=0.25t$, $t_2=0.8t$, and $R=8000$: we show the numerical and analytical scaling of the integral in Eq.~\eqref{eq:in-stiffness_integral}. The numerical scaling of the diagonal components $D_{00}=D_{11}$ agrees well with the analytical formula up to the addition of a constant independent of $L$.}
\label{fig:stiffness_scaling}
\end{figure*}

We have checked our results for $D_{s,\mathrm{geom}}$ also numerically (see Fig.~\ref{fig:stiffness_scaling}). In the numerical calculation we adopt the essence of a method for calculating the Berry curvature in a discretized Brillouin zone~\cite{Fukui2005} to efficiently compute the quantum geometric tensor $G_{ij}$ and utilize Eq.~\eqref{eq:gij_Gij}.
We find that the numerical and analytical results agree well up to the addition of a constant independent of $L$. We attribute this constant to the drumhead surface states coexisting with the zeroth PLL bulk states. 

\section{Strain implementation \label{App:StrainImplementation}}

We follow Ref.~\onlinecite{Shapourian2015} for the implementation of strain into a tight-binding model.
The tight-binding parameters $t$ represent orbital overlaps which are modified as the sample is strained because the bond lengths are changed.
To linear order, this correction can be expressed as follows,
\begin{equation}
t(\mathbf{r}_0 + \delta\mathbf{r}) = t(\mathbf{r}_0) + \frac{(\mathbf{r}_0\cdot\delta\mathbf{r})}{r_0}
\frac{\partial t}{\partial r}\bigg|_{\mathbf{r}_0},
\end{equation}
where $\mathbf{r}_0$ is the hopping vector between the two orbitals involved in equilibrium, $\delta\mathbf{r}$ is the deviation from the equilibrium value $\mathbf{r}_0$, and $t(\mathbf{r})$ is the hopping function. In general, there can be other correction terms due to the change of the angle between unlike orbitals, but we assume here that these contributions are sufficiently small.
By writing $\delta\mathbf{r}= U\mathbf{r}_0$, with the strain tensor $U=(u_{ij})$, as well as $\mathbf{r}_0 = r_0\mathbf{e}_{0}$ and $t(\mathbf{r}_0)=t_0$, the equation above becomes,
\begin{equation}
t(\mathbf{r}_0 + \delta\mathbf{r}) = t_0 
+ \left( \mathbf{e}_{0}\cdot U\mathbf{e}_{0} \right)r_0
\frac{\partial t}{\partial r}\bigg|_{\mathbf{r}_0}.
\end{equation}
For simplicity, we further assume that the hopping function can be approximated by $t(r) = t_0r_0/r$, such that we end up with the following expression:
\begin{equation}
t(\mathbf{r}) = t_0\,\left( 1 - \mathbf{e}_{0}\cdot U\mathbf{e}_{0} \right).
\label{eq:hopping_strain} 
\end{equation}
In the numerical strain implementation for CaAgP and rhombohedral graphite, all hopping terms were modified according to this formula, assuming that the dominant strain contribution comes from the $u_{11}$ component for a cylindrical substrate bent in the $x$ direction.

\section{Tight-binding model for rhombohedral graphite \label{App:Graphite}}

Our calculations for rhombohedral graphite in the main text are based on the following model~\cite{Hyart2018},
\begin{equation}
H(\mathbf{k}) = 
\begin{pmatrix}
\Theta(\mathbf{k}) & \Phi(\mathbf{k})\\
\Phi^*(\mathbf{k}) & \Theta(\mathbf{k})
\end{pmatrix}
\label{eq:graphite_ham_1}
\end{equation}
with
\begin{eqnarray}
\Phi(\mathbf{k}) &=& -\gamma_0 \sum_i e^{i\boldsymbol{\delta}_i\cdot\mathbf{k}} 
- \gamma_1\, e^{ibk_z}
- \gamma_3\, e^{-ibk_z}\sum_i e^{-i\boldsymbol{\delta}_i\cdot\mathbf{k}} \\
\Theta(\mathbf{k}) &=& -\gamma_2 \sum_i e^{i\mathbf{n}_i\cdot\mathbf{k}}
-\gamma_4\bigg(
e^{ibk_z}\sum_i e^{-i\boldsymbol{\delta}_i\cdot\mathbf{k}} 
+ e^{-ibk_z}\sum_i  e^{i\boldsymbol{\delta}_i\cdot\mathbf{k}} 
\bigg).
\end{eqnarray}
Here, $b$ is the distance between adjacent layers, $\boldsymbol{\delta}_i$ are the intra-layer nearest-neighbor hopping vectors, and $\mathbf{n}_i$ are the intra-layer next-nearest-neighbor hopping vectors~\cite{CastroNeto2009}.
We take the model parameters from literature~\cite{Ijas2013}: $\gamma_0=2.58\,\mathrm{eV}$, $\gamma_1=0.34\,\mathrm{eV}$, $\gamma_2=0$, $\gamma_3=0.17\,\mathrm{eV}$, and $\gamma_4=0.04\,\mathrm{eV}$.

For the implementation of the cylindrical strain profile, we assume that the bending direction is along the  $x$ direction. This modifies the hopping amplitudes of our tight-binding model according to Eq.~\eqref{eq:hopping_strain}.
Hence,
\begin{eqnarray}
\Phi(\mathbf{k})\rightarrow\Phi(\mathbf{k}) &=& -\gamma_0 \sum_i \bigg[1 - u_{11} \Big(\frac{\boldsymbol{\delta}_i}{|\boldsymbol{\delta}_i|} \cdot \mathbf{e}_1\Big)^2\bigg]\, e^{i\boldsymbol{\delta}_i\cdot\mathbf{k}} 
- \gamma_1\, e^{ibk_z}
- \gamma_3\, e^{-ibk_z}\sum_i \bigg[1 - u_{11} \frac{(\boldsymbol{\delta}_i \cdot \mathbf{e}_1)^2}{|\boldsymbol{\delta}_i|^2 + b^2}\bigg]\, e^{-i\boldsymbol{\delta}_i\cdot\mathbf{k}}
\label{eq:graphite_strain} 
\end{eqnarray}
and
\begin{eqnarray}
\Theta(\mathbf{k}) \rightarrow \Theta(\mathbf{k}) &=& -\gamma_2 \sum_i \bigg[1 - u_{11} \Big(\frac{\mathbf{n}_i}{|\mathbf{n}_i|} \cdot \mathbf{e}_1\Big)^2\bigg]\, e^{i\mathbf{n}_i\cdot\mathbf{k}}\nonumber\\
&&{} -\gamma_4\Bigg(
e^{ibk_z}\sum_i \bigg[1 - u_{11} \frac{(\boldsymbol{\delta}_i \cdot \mathbf{e}_1)^2}{|\boldsymbol{\delta}_i|^2 + b^2}\bigg]\, e^{-i\boldsymbol{\delta}_i\cdot\mathbf{k}} 
+ e^{-ibk_z}\sum_i \bigg[1 - u_{11} \frac{(\boldsymbol{\delta}_i \cdot \mathbf{e}_1)^2}{|\boldsymbol{\delta}_i|^2 + b^2}\bigg]\, e^{i\boldsymbol{\delta}_i\cdot\mathbf{k}} 
\Bigg)
\end{eqnarray}

\section{Analytical solutions for the zeroth PLL wave functions in rhombohedral graphite \label{app:graphite_analytics}}

\begin{figure*}
\includegraphics[width=0.8\linewidth]{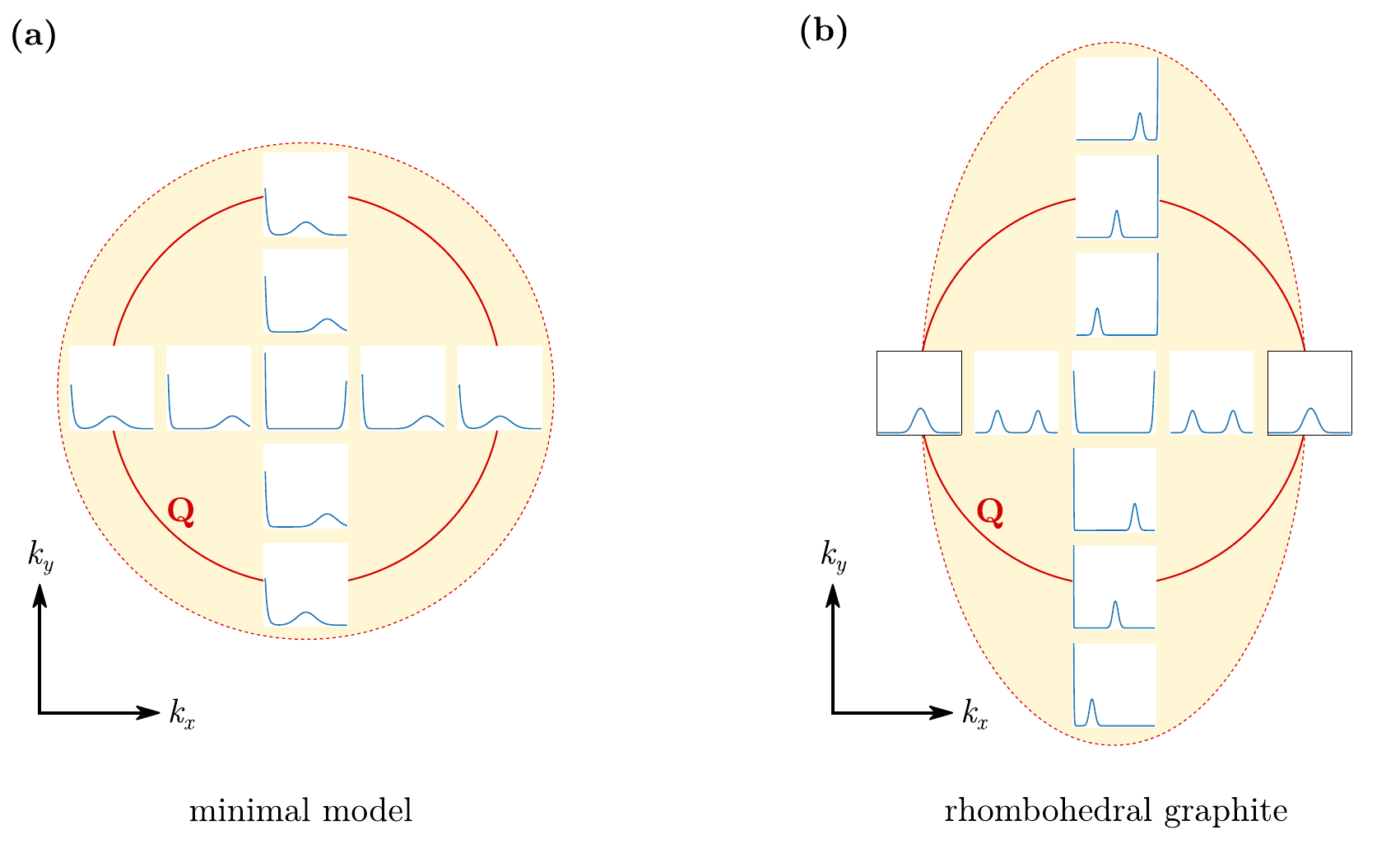}
\caption{Schematic of the zeroth PLL wave functions across the flat band. The red bold line is the nodal loop in the unstrained system parametrized by a vector $\mathbf{Q}$, whereas the red dashed line indicates the edge of the flat band (beige area) in the strained system. The small panels show the the absolute value $|\psi(z)|$ of the wave functions at various momenta within the flat band. (a) Minimal model around $\Gamma$: the shifts of the bulk wave functions are isotropic with respect to the center of the flat band. The flat band is a circular disk with radius larger than $|Q|$. (b) Rhombohedral graphite around $K$: the flat band is elongated in the $k_y$ direction and forms an elliptical disk. The structure and the behavior of the PLL wave functions are anisotropic.}
\label{fig:wf_shifts}
\end{figure*}

From our numerical analysis, we find that the structure and shift behavior of the zeroth PLL wave functions differ from those of our minimal model discussed in the main text (see Fig.~\ref{fig:wf_shifts} for a comparison). In particular, we observe two PLL bulk states along one momentum-space direction, whereas a bottom/top surface state coexist with a PLL bulk state along the perpendicular direction. Furthermore, the zeroth PLL is elongated in only one direction in contrast to the isotropic growth in the minimal model. In this section, we want to shed more light onto these findings by deriving analytical solutions for the zeroth PLL wave functions.

We start from the Hamiltonian defined in Eq.~\eqref{eq:graphite_ham_1}. 
To proceed analytically, we set $\gamma_2=\gamma_3=\gamma_4=0$, such that $\Theta(\mathbf{k})=0$.
By using in $\Phi(\mathbf{k})$ the intra-layer nearest-neighbor hopping vectors
\begin{eqnarray}
\boldsymbol{\delta}_1 &=& \frac{a}{2}(1, \sqrt{3},0) \equiv (\delta_x, \delta_y,0),\\
\boldsymbol{\delta}_2 &=& \frac{a}{2}(1, -\sqrt{3},0) = (\delta_x, -\delta_y,0),\\
\boldsymbol{\delta}_3 &=& -a\,(1,0,0),
\end{eqnarray}
and after implementing the strain terms according to Eq.~\eqref{eq:graphite_strain},
we obtain
\begin{eqnarray}
\Phi(\mathbf{k}) &=& -2\gamma_0 \left(1-\frac{u_{11}}{4}\right) \cos{(\delta_y k_y)}\,e^{i\delta_x k_x}
- \gamma_0\, (1-u_{11}) e^{-i a k_x}
- \gamma_1\, e^{ibk_z},
\label{eq:strain_Hamiltonian_graphite}
\end{eqnarray}
with $u_{11}=z/R$.

We now expand the strain Hamiltonian in Eq.~\eqref{eq:strain_Hamiltonian_graphite} around the $\mathbf{K}$ ($\mathbf{K}'$) point with $\mathbf{k} = \mathbf{K}^{(')} + \mathbf{Q}$.
This leads to
\begin{eqnarray}
\Phi_{\mathbf{K}^{(')}}  &=& 
-\frac{3a}{2} \gamma_0\, e^{-i2\pi/3}(\pm Q_y - iQ_x)\nonumber\\
&&{} + \frac{3}{4} \gamma_0\, e^{-i2\pi/3}\, u_{11} \left(1 \pm\frac{a}{2} Q_y - i\frac{3a}{2} Q_x\right)\nonumber\\
&&{}- \gamma_1\, e^{ibk_z}.
\label{eq:strain_Hamiltonian_graphiteK}
\end{eqnarray}
Note that we have only expanded in the $k_x$ and $k_y$ directions, since the nodal line extends over the whole BZ in the $k_z$ direction. Before we proceed, we first look into the structure of these nodal lines.

\subsection{Nodal lines of the unstrained system}

The nodal lines of the unstrained system ($u_{11}=0$) spiral around axes parallel to the $k_z$ axis that go through the $\mathrm{K}$ and $\mathrm{K}'$ points of the hexagonal BZ,
\begin{eqnarray}
\mathbf{K} &=& \frac{2\pi}{3\sqrt{3}a} (\sqrt{3},1,0) \equiv (K_x,K_y,0),\\
\mathbf{K}' &=&\frac{2\pi}{3\sqrt{3}a} (\sqrt{3},-1,0) = (K_x,-K_y,0).\\
\end{eqnarray}
Close to these axes, the energies of the Hamiltonian (with $u_{11}=0)$ at momenta $\mathbf{k}=\mathbf{K}^{(')} + \mathbf{Q}$ are to first order in $Q_x$ and $Q_y$
\begin{eqnarray}
E^2 &=& |\Phi(\mathbf{k})|^2 = \frac{9}{4}\gamma_0^2 a^2 (Q_x^2 + Q_y^2) +\gamma_1^2 \nonumber\\
{}&&- 3\gamma_0 \gamma_1 \left[ \sin{(aK_x + bk_z)}\, aQ_x
\mp \cos{(aK_x + bk_z)}\, aQ_y \right],
\end{eqnarray}
which leads to the following equation,
\begin{eqnarray}
\frac{4E^2}{9\gamma_0^2} &=& \left[aQ_x - \frac{2\gamma_1}{3\gamma_0}\sin{(aK_x + bQ_z)}\right]^2
+ \left[ aQ_y \pm \frac{2\gamma_1}{3\gamma_0}\cos{(aK_x + bQ_z)}\right]^2.
\end{eqnarray}
Hence, the zero-energy states lie on a spiral given by
\begin{eqnarray}
aQ_{0,x} &=& \frac{2\gamma_1}{3\gamma_0}\sin{(2\pi/3 + bQ_{0,z})},\\
aQ_{0,y} &=& \mp \frac{2\gamma_1}{3\gamma_0}\cos{(2\pi/3 + bQ_{0,z})}.
\end{eqnarray} 
This is a crucial difference to our minimal model, where the nodal line was confined to the plane $k_z=0$.

\subsection{Analytical solutions in the $k_y$ direction with respect to $\mathbf{K}$}

We now look at a particular $\mathbf{k}$-space subset around $\mathbf{K}$, for which $Q_x=0$. 
As for our minimal model, we expand $\Phi_{\mathbf{K}}$ around a nodal point $\mathbf{Q}_0=(0,Q_{0,y},Q_{0,z})$ to leading order in momentum and strain. 
We obtain
\begin{eqnarray}
\Phi(\mathbf{q}) &=&
\frac{3}{4}\gamma_0 \, e^{-i2\pi/3}\, u_{11} \left(1 + \frac{a}{2}Q_{0,y} \right)
 - \frac{3a}{2}\gamma_0 \, e^{-i2\pi/3} q_y
- i\gamma_1\, e^{ibQ_{0,z}}\,b q_z,
\end{eqnarray}
where we have used $\mathbf{q} = \mathbf{Q} - \mathbf{Q}_{0}$.
We further have that
\begin{eqnarray}
e^{i(bQ_{0,z} + 2\pi/3)} &=& \cos{(bQ_{0,z} + 2\pi/3)} + i\sin{(bQ_{0,z} + 2\pi/3)}
= -\frac{3a}{2}\frac{\gamma_0}{\gamma_1}\, Q_{0,y}.
\end{eqnarray}
Hence, after a change of basis we end up with the following expression
\begin{eqnarray}
\Phi(\mathbf{q}) &=&
\frac{3}{4}\gamma_0 \, u_{11} \left(1 + \frac{a}{2}Q_{0,y}\right)
- \frac{3a}{2}\gamma_0 q_y 
+ i\frac{3a}{2}\gamma_0 Q_{0,y} \,b q_z.
\end{eqnarray}
To obtain the zero-energy solutions of the expanded strain Hamiltonian, we have to solve
\begin{eqnarray}
\begin{pmatrix}
0 & \Phi(\mathbf{q})\\
\Phi^*(\mathbf{q}) & 0
\end{pmatrix}
\begin{pmatrix}
\psi_1\\
\psi_2
\end{pmatrix}
= 0.
\end{eqnarray}
After a change of basis and replacing $iq_z\rightarrow \partial_z$, this leads us to the following set of equations around $Q_{0,y} = \pm 2\gamma_1/3a\gamma_0 \equiv \pm Q$: 
\begin{eqnarray}
\bigg[-\frac{z}{R}\left(\gamma_1 \pm 3\gamma_0\right) 
+ 6a\gamma_0 q  - 4b \gamma_1 \partial_z \bigg]\, \psi_1 &=& 0,\\
\bigg[+\frac{z}{R} \left(\gamma_1 \pm 3\gamma_0\right) 
- 6a\gamma_0 q - 4b \gamma_1 \partial_z \bigg]\, \psi_2 &=& 0,
\end{eqnarray}
where we have set $q_y\equiv q$ for $Q_{0,y}>0$, and  $q_y\equiv -q$ for $Q_{0,y}<0$.
With this definition, momenta with $q<0$ ($q>0$) are inside (outside) the projected nodal circle. For sufficiently small nodal lines, we can assume that $\gamma_1<\gamma_0$. 
In this case, the sign in front of $\gamma_0$ determines the overall sign of the terms $\gamma_1 \pm 3\gamma_0$.
We therefore define,
\begin{equation}
\sigma_\pm = \sqrt{\frac{4Rb\gamma_1}{(3\gamma_0 \pm\gamma_1)}}
\end{equation}
and 
\begin{equation}
\lambda = \frac{3a\gamma_0}{2b\gamma_1} q.
\end{equation}
With this, we obtain the following zero-energy solutions:
\begin{itemize}
\item for $Q_{0,y}>0$,
\begin{eqnarray}
\psi_1(z) &=& A_1\, e^{-z^2/2\sigma_+^2}\, e^{\lambda z}
\propto e^{-\left(z-\lambda\sigma_+^2\right)^2 / 2\sigma_+^2},\\
\psi_2(z) &=& A_2\, e^{+z^2/2\sigma_+^2}\, e^{-\lambda z}
\propto e^{+\left(z-\lambda\sigma_+^2\right)^2 / 2\sigma_+^2},
\end{eqnarray}
\item for $Q_{0,y}<0$,
\begin{eqnarray}
\psi_1(z) &=& B_1\, e^{+z^2/2\sigma_-^2}\, e^{\lambda z}
\propto e^{+\left(z+\lambda\sigma_-^2\right)^2 / 2\sigma_-^2},\\
\psi_2(z) &=& B_2\, e^{-z^2/2\sigma_-^2}\, e^{-\lambda z}
\propto e^{-\left(z+\lambda\sigma_-^2\right)^2 / 2\sigma_-^2}.
\end{eqnarray}
\end{itemize}
Here, the solutions describe a PLL bulk state and a top-surface state for $Q_{0,y}>0$, but a PLL bulk state and a bottom-surface state for $Q_{0,y}<0$. Furthermore, the PLL bulk-state shifts along this direction are not symmetric with respect to the center of the flat band: starting from $Q_{0,y}<0$, the bulk-state shifts towards the top surface as we move towards the center of the flat band, whereas it shifts towards the bottom surface, if we do the same starting from $Q_{0,y}>0$. This is in qualitative agreement with the numerical results [see Fig.~\ref{fig:wf_shifts}(b)] and we have also confirmed that analytical and numerical solutions agree quantitatively.

\subsection{Analytical solutions in the $k_x$ direction with respect to $\mathbf{K}$}

Let us now turn to the perpendicular subset in momentum space for which $Q_y=0$. Before we proceed, recall from Fig.~\ref{fig:wf_shifts}(b) that the flat band does not grow in the $Q_x$ direction. In particular, we know from numerics that the two PLL bulk states hybridize at the nodal-line momenta $Q_{x,0}$. Therefore, an expansion around a nodal point as before is not a promising approach to find zero-energy solutions.

Instead, we will work with the initial expansion of the strain Hamiltonian around $\mathbf{K}$ from Eq.~\eqref{eq:strain_Hamiltonian_graphiteK}, with $\mathbf{k} = \mathbf{K} + \mathbf{q}$ and $\mathbf{q} = (q_x,0,0)$,
\begin{eqnarray}
\Phi(q_x,k_z) &=& 
\frac{3a}{2}\gamma_0\, e^{-i2\pi/3}\,iq_x
+ \frac{3}{4} \gamma_0\, e^{-i2\pi/3}\, u_{11} \left(1  - \frac{3a}{2}\, i q_x\right)
- \gamma_1\, e^{ibk_z}.
\end{eqnarray}
Note that we have, for notational consistency, renamed $\mathbf{Q}\rightarrow\mathbf{q}$.
After a change of basis, this becomes
\begin{eqnarray}
\Phi(q_x,k_z)  &=& 
\frac{3a}{2}\gamma_0\, iq_x
+ \frac{3}{4} \gamma_0\, u_{11} \left(1  - \frac{3a}{2}\, i q_x\right)
- \gamma_1\, e^{i(bk_z+2\pi/3)}.
\label{eq:strain_Hamiltonian_graphiteK_qx}
\end{eqnarray}

Recall that we have not yet expanded the Hamiltonian along $k_z$.
From numerics, we obtain that the Fourier-transformed PLL bulk solutions are centered at $k_z$ values that depend on $q_x$, suggesting that the corresponding expansion point along the $k_z$ direction should be a function of $q_x$.
In particular, we find the following relation between the bulk-state centers $K_z$ and $q_x$
\begin{eqnarray}
q_x(K_z) = Q\cos{\left(bK_z + \pi/6\right)},
\end{eqnarray}
where $Q=2\gamma_1/3a\gamma_0$ is the radius of the nodal spiral projected into the $k_x$-$k_y$ plane.
For fixed $q_x$, there are two centers $K_{z,1}$ and $K_{z,2}$ corresponding to the two PLL bulk states,
\begin{eqnarray}
b K_{z,1} &=& \arccos{\left(\frac{q_x}{Q}\right)} - \pi/6 \equiv K_{z,+}\\
b K_{z,2} &=& -\arccos{\left(\frac{q_x}{Q}\right)} - \pi/6 \equiv K_{z,-}
\end{eqnarray}
We now expand Eq.~\eqref{eq:strain_Hamiltonian_graphiteK_qx}  along the $k_z$ direction around each of these points separately.
We obtain 
\begin{eqnarray}
\Phi(\mathbf{q}) &=& 
\frac{3}{4} \gamma_0\, u_{11} \left(1  - \frac{3a}{2}\, i q_x\right)
\pm \frac{3a}{2}\gamma_0 \sqrt{Q^2-q_x^2}
- \frac{3a}{2}\gamma_0\left(\mp \sqrt{Q^2-q_x^2} + iq_x\right) ibq_z
\end{eqnarray}
We are interested in zero-energy solutions, i.e., in the solutions of 
\begin{eqnarray}
\begin{pmatrix}
0 & \Phi\\
\Phi^* & 0
\end{pmatrix}
\begin{pmatrix}
\psi_1\\
\psi_2
\end{pmatrix}
= 0.
\end{eqnarray}
For notational simplicity, we set $a=b=1$ in the following.
After a change of basis and replacing $iq_z\rightarrow \partial_z$, we obtain the following set of equations
\begin{eqnarray}
\Bigg[\frac{z}{2RQ^2}\left(\mp\sqrt{Q^2-q_x^2} - iq_x\right)\left(1 - \frac{3}{2}\,iq_x\right)
- \frac{1}{Q^2}\left( Q^2 - q_x^2 \pm iq_x \sqrt{Q^2-q_x^2}\right)
-\partial_z\Bigg]\psi_1 &=& 0\\
\Bigg[\frac{z}{2RQ^2}\left(\pm\sqrt{Q^2-q_x^2} - iq_x\right)\left(1 + \frac{3}{2}\,iq_x\right)
+ \frac{1}{Q^2}\left( Q^2 - q_x^2 \mp iq_x \sqrt{Q^2-q_x^2}\right)
-\partial_z\Bigg]\psi_2 &=& 0
\end{eqnarray}
To analyze the solutions of these equations, we introduce the following short-hand notations:
\begin{eqnarray}
\lambda_\pm &=& \frac{1}{Q^2}\left( Q^2 - q_x^2 \pm iq_x \sqrt{Q^2-q_x^2}\right),\\
\omega_\pm &=&\frac{1}{2RQ^2}\left(\pm\sqrt{Q^2-q_x^2} + iq_x\right)\left(1 - \frac{3}{2}\,iq_x\right).
\end{eqnarray}
With this, the set of equations from above becomes
\begin{eqnarray}
\frac{d\psi_1}{dz} &=& -\lambda_\pm\psi_1 - \omega_\pm z\, \psi_1\\
\frac{d\psi_2}{dz} &=& +\lambda_\pm^* \psi_2 + \omega_\pm^* z\, \psi_2\\
\end{eqnarray}
The solutions of these equations are
\begin{eqnarray}
\psi_{1,\pm}(z) &=& A_\pm e^{-\lambda_\pm z} e^{-\frac{1}{2}\omega_\pm z^2}\\
\psi_{2,\pm}(z) &=& B_\pm e^{\lambda^*_\pm z} e^{+\frac{1}{2}\omega^*_\pm z^2}.
\end{eqnarray}
Let us look at the absolute values of these solutions,
\begin{eqnarray}
|\psi_{1,\pm}(z)|^2 &=& |A_\pm|^2 e^{-2\mathrm{Re}(\lambda_\pm) z} e^{-\mathrm{Re}(\omega_\pm) z^2}\nonumber\\
&=& |A_\pm|^2 \exp{\left(-2\left[1-(q_x/Q)^2\right] z\right)} 
\exp{\left[ -\frac{1}{2RQ^2} \left( \pm\sqrt{Q^2-q_x^2} + 3q_x^2/2 \right)z^2\right]}, 
\end{eqnarray}
and
\begin{eqnarray}
|\psi_{2,\pm}(z)|^2 &=& 
|B_\pm|^2 \exp{\left(+2\left[1-(q_x/Q)^2\right] z\right)} 
\exp{\left[ +\frac{1}{2RQ^2} \left( \pm\sqrt{Q^2-q_x^2} + 3q_x^2/2 \right)z^2\right]}. 
\end{eqnarray}

In the limit $q_x=0$, this simplifies to:
\begin{eqnarray}
|\psi_{1,\pm}(z)|^2 &=& 
|A_\pm|^2 \exp{\left(-2 z\right)} 
\exp{\left[ \mp\frac{1}{2RQ}\,z^2\right]} 
= |\tilde{A}_\pm|^2 \exp{\left[ \mp\frac{1}{2RQ}\left(z \pm 2RQ\right)^2\right]} \\
|\psi_{2,\pm}(z)|^2 &=& 
|B_\pm|^2 \exp{\left(+2 z\right)} 
\exp{\left[ \pm\frac{1}{2RQ}\,z^2\right]}
=|\tilde{B}_\pm|^2 \exp{\left[ \pm\frac{1}{2RQ}\left(z \pm 2RQ\right)^2\right]} . 
\end{eqnarray}
Hence, the Gaussian solutions are $\psi_{1,+}$ and $\psi_{2,-}$ with centers at $\pm 2RQ$.

\begin{figure*}
\includegraphics[width=0.85\linewidth]{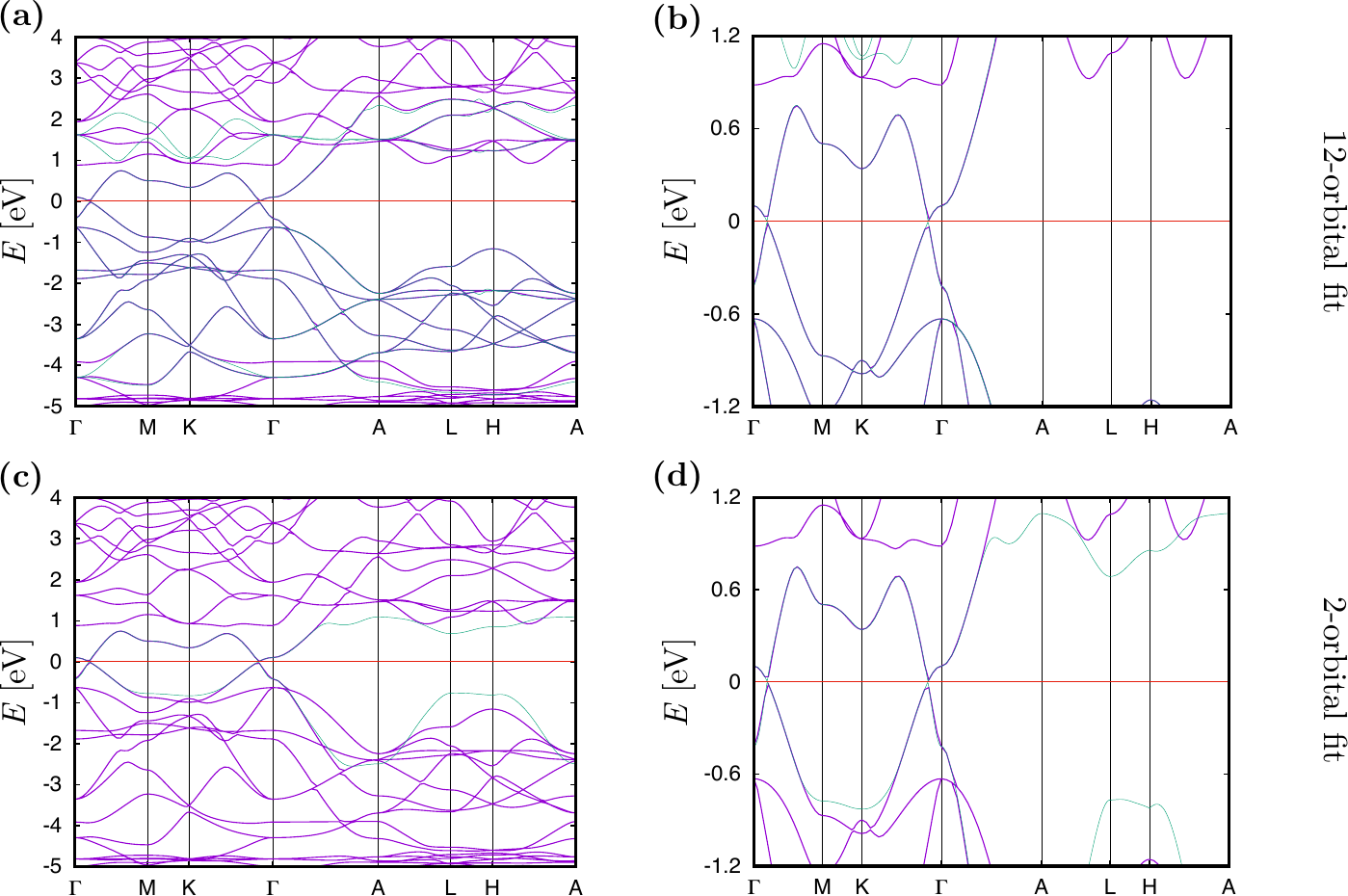}
\caption{DFT band structure (purple) and bands obtained from Wannier fits (cyan) for CaAgP: the first (second) row shows 12 (two) interpolated Ag-s and
P-p bands obtained using the Wannier functions for CaAgP. 
The Fermi level is set to zero.
In (b) and (d), we present a magnified plot of the bands shown in (a) and (c), respectively.}
\label{fig:DFT_CaAgP}
\end{figure*}

For general $q_x$ with $|q_x|<Q$, $\psi_{1,+}$ is always a Gaussian because the coefficient in parentheses in front of $z^2$ is always positive.
This behavior is different for $\psi_{-}$, for which this coefficient changes sign at some critical $q_{x,c}$ with $q_{x,c}^2   = \frac{2}{9}\left(\sqrt{9Q^2+1}-1\right)$.
In particular, at $q_x=\pm Q$ we get
\begin{eqnarray}
|\psi_{1,\pm}(z)|^2 &=& 
|A_\pm|^2
\exp{\left[ -\frac{3}{4R}\,z^2\right]} \\
|\psi_{2,\pm}(z)|^2 &=& 
|B_\pm|^2
\exp{\left[ +\frac{3}{4R}\,z^2\right]}, 
\end{eqnarray}
such that, in this case, $\psi_{1,+}$ and $\psi_{2,+}$ are the bulk PLL solutions with centers at $z=0$.
For sufficiently small $Q$, nevertheless, we have that $|q_{x,c}|\approx Q$ such that $\psi_{2,-}$ describes the second Gaussian solution for nearly the whole interior of the flat band along the $q_x$ direction. We have checked that these solutions agree well with the numerically obtained solutions shown in Fig.~\ref{fig:wf_shifts}(b). The other two analytical solutions $\psi_{1,-}$ and $\psi_{2,+}$ are, however, not obtained in numerics implying that they are not valid approximate solutions of the full Hamiltonian.  

To summarize, the valid solutions along the $q_x$ direction are:
\begin{eqnarray}
\psi_{1,+}(z) &\propto& e^{-\lambda_+ z}\, e^{-\frac{1}{2}\omega_- z^2}\\
\psi_{2,-}(z) &\propto& e^{+\lambda^*_- z}\, e^{+\frac{1}{2}\omega^*_- z^2}.
\end{eqnarray}

\section{DFT calculations for C\lowercase{a}A\lowercase{g}P \label{App:DFT_CaAgP}}

In this section, we provide details on the construction of the tight-binding model for CaAgP used in the main text.

We have performed first-principles density functional theory (DFT)
calculations by using the VASP package based
on the plane-wave basis set and projector augmented wave method~\cite{Kresse:1996_CMS,Kresse:1999_PRB}.
A plane-wave energy cut-off of $270 \, \mathrm{eV}$ has been used.
For the treatment of exchange correlations, the Perdew-Burke-Ernzerhof
generalized gradient approximation (GGA)~\cite{Perdew:1996_PRL} has been applied. 
We have used a $12 \times 12 \times 18$ k-point grid centered at $\Gamma$. 
After computing the Bloch wave functions in DFT, we construct corresponding Wannier functions (WFs)~\cite{Marzari:1997_PRB,Souza:2001_PRB}
using the WANNIER90 code~\cite{Mostofi:2008_CPC}.
To extract the orbital character of the
electronic bands at low energies, we use the Slater-Koster interpolation scheme based on the WFs.
Furthermore, we neglect spin-orbit interactions, which are small in this material.

As a first step, we construct a 12-orbital model based on the $3p$ orbitals of the three P atoms and on the $5s$ orbitals of the three Ag atoms in the unit cell.
Our band structure results are shown in Figs.~\ref{fig:DFT_CaAgP}(a) and~(b) and are in agreement with the literature~\cite{Yamakage:2016_JPSJ}.
The match between the DFT band structure and the interpolated band structure 
obtained from the WFs is good around the Fermi level.
To obtain a simpler model catching the essential physics only at low energies, we next construct an effective two-orbital tight-binding model based on
one $p_z$ orbital centered at one P atom and on one s-orbital centered at the middle of a triangle of Ag atoms.
The band interpolation of the two-orbital model is shown in Figs.~\ref{fig:DFT_CaAgP}(c) and~(d).
As we can see from the comparison between Figs.~\ref{fig:DFT_CaAgP}(b) and~(d), we do not lose accuracy between -0.60 and 0.85 eV 
moving from the 12-orbital to the two-orbital model.

This model, as obtained from the Wannier interpolation, still has a large number of parameters. To reduce this number to a managable value, we finally set an energy cut-off of $6\,\mathrm{meV}$ for the tight-binding parameters such that the dispersion close to the Fermi level is still captured correctly. 
This results in a two-band model with 37 different parameters, which is used for the strain implementation in the main text.

\section{Effect of orbital-mixing strain terms}
\label{App:orbital_mixing_strain}

\begin{figure}[h]
\includegraphics[width=0.9\linewidth]{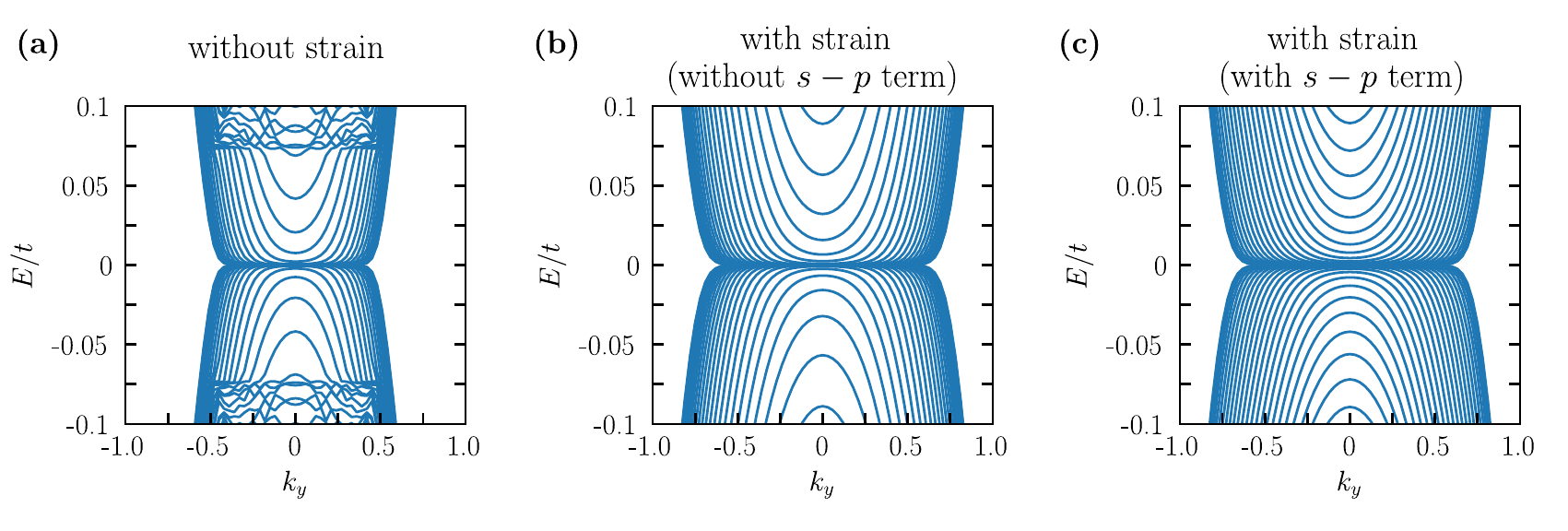}
\caption{Energy subbands close to $E=0$ for the minimal model with open boundary conditions in both $x$ and $z$ direction, and model parameters $t_1=0.25t$, $t_2=0.8t$, $L_x=L_z=100$. (a) Unstrained system. (b) Strained system without orbital-mixing term.  (c) Strained system with orbital-mixing $s-p$ terms. Here, we have chosen $R=200$.}
\label{fig:additional_sp_term}
\end{figure}

In our considerations, we have disregarded the effect of non-universal strain terms that might arise due to the specific orbital structure of the model, such as strain-enabled hopping terms between unlike orbitals.
In this section, we show for our minimal model that including these terms does not change the results significantly.

In Sec.~\ref{app:pseudo_field}, we implemented the cylindrical strain profile focusing on the change in the $x$-bond length, which led to the strain Hamiltonian $H_0(\mathbf{k})$ shown in Eq.~\eqref{eq:H_0}. These strain terms are present independent of the orbital structure.
If the two orbital degrees of freedom correspond to specific $s$ and $p_z$ orbitals, the considered strain profile also makes $s-p$ hopping available in the $x$ direction. There exists several possible approximation schemes to account for the mixing of the $s$ and $p$ orbitals (and their validity depends on the details of the materials), but in this section we follow the specific approach proposed in Ref.~\onlinecite{Shapourian2015} for simplicity. This way, we determine the mixing term to be $2t_2\sigma^x\, u_{31} \sin{k_x}$, with the strain tensor component $u_{31}=u_{13}=x/2R$.
Hence, the full strain Hamiltonian reads
\begin{eqnarray}
\tilde{H}(\boldsymbol{k}) &=&
\sigma^{z}[6t-t_{1}-2t (1-u_{11}) \cos k_{x}-2 t \sum_{i=y,z} \cos k_{i}] +2t_{2}\sigma^{x} \left(
\sin k_{z} + u_{31}\,\sin k_x \right).
 \label{eq:H_1}
\end{eqnarray}
Recall that the strain terms in $H_0(\mathbf{k})$ [see Eq.~\eqref{eq:H_0}] allow to retain the in-plane translational symmetry of the system. 
Hence, both $k_x$ and $k_y$ are good quantum numbers enabling us to visualize the spectrum of the strained system with finite width $L_z$ in an effective 2D Brillouin zone (see Fig.~1 of the main text).
On the contrary, the presence of $u_{31}$ in Eq.~\eqref{eq:H_1} breaks translational symmetry also in the $x$ direction.
Hence, only $k_y$ remains a good quantum number and open boundary conditions have to be applied in the $x$ and $z$ directions~\footnote{Note that this comes at considerably greater computational cost: previously, for each momentum $(k_x, k_y)$ we had to diagonalize a $2N_z\times 2N_z$ matrix to obtain the energies and states of the system, where $N_z$ is the number of lattice sites in the $z$ direction. Now, we instead have to deal with a matrix of size $2N_zN_x\times 2N_zN_x$ for each $k_y$, with $N_x$ the number of sites in the $x$ direction.}. 

Figure~\ref{fig:additional_sp_term} shows the energy bands of the system in such a geometry, i.e., the system is finite in $x$ and $z$ with widths of $L_x$ and $L_z$, respectively, in units of the lattice constant of the unstrained system.
For comparison, in Figs.~\ref{fig:additional_sp_term}(a) and~(b) we also plot the energies of the unstrained system and the strained system without the $s-p$ terms. 
We observe a number of flat subbands close to $E=0$ comprising the top- and bottom-surface drumhead states in the case of the unstrained system, and surface and bulk pseudo-Landau level (PLL) states in the case of the strained system.
Most importantly, we find that the $k_y$ extent and the number of flat subbands are larger than in the unstrained system.
This reflects the growing density of states as strain is applied.

Finally, in Fig.~\ref{fig:additional_sp_term}(c) we plot the energy subbands of the strained system including the $s-p$ strain terms.
We find that the spectrum is strikingly similar to Fig.~\ref{fig:additional_sp_term}(b).
In particular, number and extent of the flat subbands close to $E=0$ are larger than in the unstrained system and almost identical to Fig.~\ref{fig:additional_sp_term}(b). 
The only difference is the subband level spacing away from $E=0$.
Moreover, we have checked that the flat subbands indeed consist of surface states and bulk PLL states behaving in a way identical to what we observe in the absence of the additional $s-p$ strain terms: the centers of the bulk PLL shift as we move along $k_y$ or along the subband index.

This confirms that orbital-mixing strain terms have only a minor effect on the details of the spectrum and the system still supports 3D flat bands.

\section{Portion of the Brillouin zone covered by the flat bands}

We have mostly assumed that $Q \ll 1$. Moreover, the validity of the leading order expansion of the strain requires that $R \gg L$. With these assumptions the portion of the Brillouin zone covered by the flat band is always small. However, the first assumption was done only for the sake of analytical transparency. Moreover, if we implement the effective magnetic field by varying the chemical composition so that the radius of the nodal line changes along the $z$-direction, we can also violate the second condition. In the extreme case we can have $Q \sim 1$ and $R \sim L$. In this case, the analytical approximations are no longer valid, but numerically we find 3D flat bands which cover most of the Brillouin zone (see Fig.~\ref{fig:BZcoverage}). Thus, we conclude that there are no significant fundamental limitations for the portion of the Brillouin zone covered by the flat bands.

\begin{figure}[h]
\includegraphics[width=0.9\linewidth]{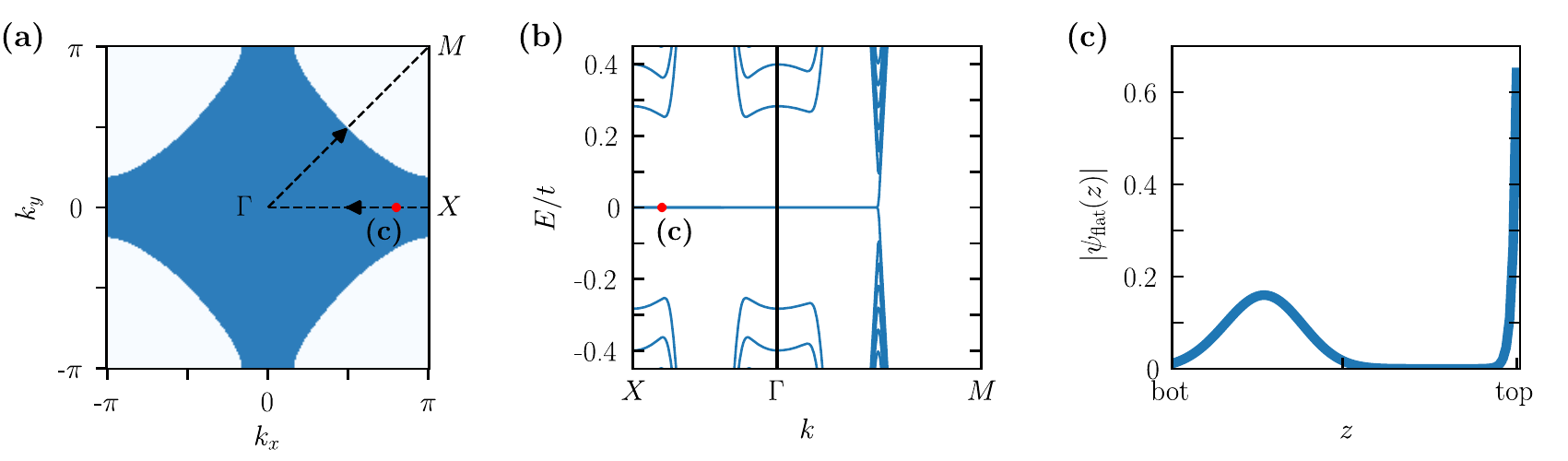}
\caption{Large flat bands in the minimal model for parameters $t_1=4t$, $t_2=t$, and $L=R=100$:
(a) extent of the flat bands (blue) at $E=0$ in the full Brillouin zone. (b) Energy bands close to $E=0$ along the path indicated in (a). (c) Wave function of the flat-band states at the momentum indicated in (a) and (b).}
\label{fig:BZcoverage}
\end{figure}

\end{widetext}

\end{document}